\documentclass[prb,twocolumn,showpacs,superscriptaddress,floatfix]{revtex4-1}

\usepackage{amsmath}
\usepackage{amssymb}
\usepackage{graphicx}
\usepackage{bm}
\usepackage{lipsum}
\usepackage[usenames,dvipsnames]{color}
\usepackage{blkarray}
\newcommand{\matindex}[1]{#1}
\usepackage{hyperref}
\usepackage[all]{hypcap}

\renewcommand{\i}{\ensuremath{\mathrm{i}}}

\begin{document}

\title{A new paradigm for the quantum spin Hall effect at high temperatures}

\author{Gang Li}
\email{ligang@shanghaitech.edu.cn}
\affiliation{School of Physical Science and Technology, ShanghaiTech University, Shanghai 200031, China}

\author{Werner Hanke}
\affiliation{Institut f\"{u}r Theoretische Physik und Astrophysik,Universit\"{a}t W\"urzburg, D-97074 W\"urzburg, Germany}

\author{Ewelina M. Hankiewicz}
\affiliation{Institut f\"{u}r Theoretische Physik und Astrophysik,Universit\"{a}t W\"urzburg, D-97074 W\"urzburg, Germany}

\author{Felix Reis}
\affiliation{Physikalisches Institut and R\"ontgen Center for Complex Material Systems, University of W\"{u}rzburg, Am Hubland, 97074 W\"{u}rzburg, Germany}

\author{J\"org Sch\"afer}
\affiliation{Physikalisches Institut and R\"ontgen Center for Complex Material Systems, University of W\"{u}rzburg, Am Hubland, 97074 W\"{u}rzburg, Germany}

\author{Ralph Claessen}
\affiliation{Physikalisches Institut and R\"ontgen Center for Complex Material Systems, University of W\"{u}rzburg, Am Hubland, 97074 W\"{u}rzburg, Germany}

\author{Congjun Wu}
\affiliation{Department of Physics, University of California, San Diego, California, 92093, USA}

\author{Ronny Thomale}
\email{rthomale@physik.uni-wuerzburg.de}
\affiliation{Institut f\"{u}r Theoretische Physik und Astrophysik,Universit\"{a}t W\"urzburg, D-97074 W\"urzburg, Germany}

\date{\today}


\begin{abstract}
The quantum spin Hall effect (QSHE) has formed the seed for contemporary research on topological quantum states of matter. Since its discovery in HgTe/CdTe quantum wells and AlGaAs/GaAs heterostructures, all such systems have so far been suffering from extremely low operating temperatures, rendering any technological application out of reach. We formulate a theoretical paradigm to accomplish the high temperature QSHE in monolayer-substrate heterostructures. Specifically, we explicate our proposal for hexagonal compounds formed by monolayers of heavy group-V elements (As, Sb, Bi) on a SiC substrate. We show how orbital filtering due to substrate hybridization, a tailored multi-orbital density of states at low energies, and large spin-orbit coupling can conspire to yield QSH states with bulk gaps of several hundreds of meV. Combined with the successful realization of Bi/SiC (0001), with a measured bulk gap of $\sim 
800$ meV reported previously [Reis {\it et al.}, 10.1126/science.aai8142 (2017)], our paradigm elevates the QSHE from an intricate quantum phenomenon at low temperatures to a scalable effect amenable to device design and engineering.
\end{abstract}
\maketitle

\section{Introduction}
\label{Sec:Intro}

Since its discovery in HgTe/CdTe quantum wells~\cite{Konig766,Roth294} and subsequent findings in AlGaAs/GaAs heterostructures,\cite{PhysRevLett.107.136603} the quantum spin Hall effect (QSHE) has become a central source of stimulation for the evolving field of topological phases in condensed matter physics.\cite{HasanKane,xlreview,Moore2010:N}
	
The QSHE, as predicted from theory,\cite{kane-05prl226801,Bernevig1757} features a two-dimensional bulk insulator whose transport properties at low energies are determined by topologically protected edge modes insensitive to elastic backscattering when the system is non-interacting or weakly interacting. Specifically, the edge is formed by one pair of counterpropagating time-reversed modes in the sense of Kramers' symmetry, which cannot be removed unless the bulk gap is closed, or, its constituting time-reversal symmetry~\cite{Wu2006:PRL,Xu2006:PRB} is explicitly broken by magnetic scatterings or spontaneously broken by strong interactions. Alternatively, the band structure of a QSH insulator is characterized by a $Z_2$ topological invariant,\cite{PhysRevLett.95.146802}taking the values $1$ if this pair of edge modes is present at a boundary termination, or $0$, if it is absent as for a trivial insulator.

The QSH boundary modes are termed helical, since their spin degree of freedom and their direction of motion are coupled. It immediately suggests a basis for spintronic device applications, where the QSH insulator could operate, for instance, as a spin splitter. In technological practice, {\it i.e.} for material realizations at room temperature, however, it becomes a challenging task to preserve the universal transport character of the QSH edge. It must hold $k_{\text{B}}T/\Delta \ll 1$, where $\Delta$ denotes the bulk gap, in order to avoid parasitic bulk conductance contributions. Note that, at the example of HgTe as the QSH material with the biggest $\Delta$ until recently, there is a plethora of different scattering sources that have to be considered\cite{PhysRevLett.110.216402} which impose themselves on the universal conductance signature of the QSH edge. In particular, due to vacancies and other types of imperfections, the achievable operational transport gap might be significantly smaller than the bulk gap.\cite{PhysRevX.3.021003} This explains why, at present, the search for wide-gap ($\gtrsim 0.2$ eV) QSH systems is a key issue in contemporary condensed matter research, and a necessary step to transcend QSHE to the realm of technological applications.

In a recent work\cite{Reis2017:S}, we have reported the experimental finding of a bismuthene-type heterostructure, {\it i.e.}, a monolayer of Bi deposited on a SiC substrate. With all evidence derived from local spectroscopy, Bi/SiC constitutes a large-gap ($\sim0.8$ eV) QSH insulator. This interpretation has been strongly supported by our successful step-by-step theoretical analysis, starting from a priori density functional theory~\cite{PhysRev.136.B864, PhysRev.140.A1133}(DFT) calculations and low-energy ``downfolding" to obtain an effective model~\cite{Reis2017:S}.

In particular, our theory emphasizes the crucial role played by an appropriate substrate (Bi honeycomb layer on SiC). The layer-substrate combination offers a kind of "best of two worlds" effect, where the substrate is not only stabilizing the quasi-2D topological insulator but, additionally, plays a pivotal role for achieving the large topological gap in a graphene-type topological system: The key result is that the large on-site, {\it i.e.} atomic spin-orbit coupling (SOC), directly determines the magnitude of the topological gap in this system, without any other small prefactor as, e.g., stemming from longer range hybridization.

Another strategy to boost the topological band gap was proposed in the $p_x$-$p_y$ honeycomb lattice systems based on the orbitally enriched Dirac cone structure \cite{ZhangGF2014}. The Bloch wave states at the Dirac points correspond to non-bonding states, {\it i.e.}, the two sublattices actually decouple at the Dirac point. In this case, the solid state gap opening is reduced to the atomic energy-level splitting problem. Hence, the atomic spin-orbit coupling can completely contribute to open the topological gaps. This study was initiated in the context of orbital-active honeycomb optical lattices with ultra-cold atoms based on the $p_x$ and $p_y$ orbitals \cite{WuCJ2007, WuCJ2008}, in which the $p_z$ orbital is pushed to high energy by imposing a strong optical confinement along the $z$-direction.
The $p_x$ and $p_y$-band structure exhibits both flat bands and dispersive bands with Dirac cones. Due to the orbital structure, they are sensitive to a topological
gap opening by applying the "shaking lattice method" to realize an orbital Zeeman term, which generates the quantum anomalous Hall state \cite{WuCJ2008a,Zhang2011:PRA}.
The Kramers doubled version of these works is the above mechanism for large gap quantum spin Hall insulators, in which the atomic spin-orbit coupling is the time-reversal invariant generalization of the orbital Zeeman term.
	
In this article, inspired by the impressive agreement for the specific example of Bi/SiC, we distill the universality aspects discovered in Ref.~\onlinecite{Reis2017:S} to elevate our theoretical concepts to a generic paradigm for a class of large-gap, and hence potentially high-temperature, QSH material compounds. Our universal effective low-energy Hamiltonian for the honeycomb monolayer reads
\begin{equation}\label{EffLowEnergyHamiltonianGeneral}
H_{\text{eff}}^{\sigma\sigma}=H_{0}^{\sigma\sigma}+\lambda_{\text{SOC}}H_{\text{SOC}}^{\sigma\sigma}+\lambda_{\text{R}}H_{\text{R}}^{\sigma\sigma},
\end{equation}
where $H_{0}^{\sigma\sigma}$, $H_{\text{SOC}}^{\sigma\sigma}$ and $H_{\text{R}}^{\sigma\sigma}$ are derived in Secs.~\ref{Sec:ElectronicStructure} and~\ref{Sec:LowEnergyModel} below. In Eq.~(\ref{EffLowEnergyHamiltonianGeneral}), starting from a spin-rotation invariant band structure contribution $H_{0}^{\sigma\sigma}$, spin-orbit coupling (SOC) comes into play via two distinct terms, $H_{\text{SOC}}^{\sigma\sigma}$ as its intrinsic atomic and $H_{\text{R}}^{\sigma\sigma}$ as its Rashba contribution related to inversion symmetry breaking in the monolayer due to the substrate. As explained below, the $\sigma$-bond states in the monolayer are sufficient to form the effective model, while the $\pi$-bonding sector enters perturbatively.
It is this $\pi$ band $p_z$-orbital sector which, due to the binding to the substrate and the resulting broken inversion symmetry perpendicular to the layer, is the driving force behind the third term in Eq.~(\ref{EffLowEnergyHamiltonianGeneral}), {\it i.e.} the Rashba SOC $\sim H_{\text{R}}^{\sigma\sigma}$. This term creates a splitting of the valence bands, which can be measured by photoemission (ARPES). This sets up an important consistency check of the topological band theory, as shown in our earlier work.\cite{Reis2017:S}

The quintessential progress implied by our effective model can be phrased as extending the Kane-Mele (KM) honeycomb model~\cite{kane-05prl226801} to a specifically chosen multi-dimensional local orbital basis. For the KM model, it is found that the bulk gap scales with the local atomic SOC of the constituent atom species forming the honeycomb monolayer, albeit in higher-order perturbation theory. It allows for a more systematic quantification of the bulk gap than in HgTe, where the bulk gap, manifests from the splitting of the inverted $\Gamma_6$ and $\Gamma_8$ bands, only appears due to the non-universal crystal symmetry breaking imposed by the CdTe quantum well geometry. Due to its single orbital nature and its two-site honeycomb unit cell, the atomic SOC term in the KM model is only allowed to appear via a second nearest-neighbor hybridization effect, which dramatically reduces the bulk gap implied by it. Instead, as elaborated on in Sec.~\ref{Sec:LowEnergyModel}, the combined honeycomb layer/substrate system, described in Eq.~(\ref{EffLowEnergyHamiltonianGeneral}) features a two-dimensional $p_x$, $p_y$ orbital basis per site, allowing atomic SOC to induce local matrix elements between the different orbitals. For $p_x$ and $p_y$, it turns out that this coupling is optimal in the sense that only the $L_z$ orbital angular momentum component couples to the spin, so there are no interference effects between different SOC matrix elements. This yields $\Delta \sim O(1) \lambda_{\text{SOC}}$, and as such bulk gaps of several hundred meV for the compounds we propose.

The article is organized as follows. In Sec.~\ref{Sec:ElectronicStructure}, the effective electronic structure, first without including spin-orbit effects, {\it i.e.} $H_{0}^{\sigma\sigma}$ as contained in Eq.~(\ref{EffLowEnergyHamiltonianGeneral}), is systematically derived and explicated, starting from the ab-initio electronic structure of (As,Sb,Bi)/SiC(0001). As one key insight from this analysis, the SiC substrate hybridization, and its lattice mismatch, is pivotal to accomplish the effective $p_x$, $p_y$ low energy electronic structure in the monolayer.

Section~\ref{Sec:LowEnergyModel} continues by including SOC to the electronic structure. As it turns out, in addition to a strong atomic SOC term determining the imposed bulk gap, the SiC substrate triggers a significant Rashba term $H_{\text{R}}^{\sigma\sigma}$ in Eq.~(\ref{EffLowEnergyHamiltonianGeneral}) which, despite not being directly responsible for the formation of the QSH state --- as mentioned above --- gives rise to characteristic modifications of the topological band structure amenable to {\it e.g.} ARPES. Furthermore, $H_{\text{R}}^{\sigma\sigma}$ might be interesting in its own right towards directions such as valleytronics. We employ a band-structure analysis to successively derive the different SOC terms in Eq.~(\ref{EffLowEnergyHamiltonianGeneral}), as we first include atomic SOC $H_{\text{SOC}}^{\sigma\sigma}$ which only acts within the $\sigma$ bond sector, and then perturbatively add the $\pi$ bond sector to include the Rashba SOC contribution $H_{\text{R}}^{\sigma\sigma}$ to its leading order in the $\sigma$ bond sector.

Section~\ref{Sec:FourBandModel} presents then a further simplified version of the 8-band (effective $\sigma$-bands) electronic structure, as derived in Sec.~\ref{Sec:LowEnergyModel}, which highlights the interplay between lattice symmetry and the SOC. By considering only states close to zero energy around the K- and K'-points, their point-group symmetries and their interplay with SOC are most naturally exploited using the circularly polarized orbital states $p_\pm=(p_x\pm\i p_y)/\sqrt{2}$, which carry definite onsite orbital angular momentum.

In Sec.~\ref{Sec:Summary}, we summarize the essential insights derived from our analysis towards to the goal of high temperature QSHE, and outline potential directions emerging from this work. Among them, we highlight the possibility of quantum anomalous Hall states seeded from our model, where the magnetic proximity effect, {\it e.g.} imposed by a suitably chosen substrate or magnetic capping layer, provides a mechanism to realize such states of matter.

\begin{figure*}[htbp]
\centering
\includegraphics[width=\linewidth]{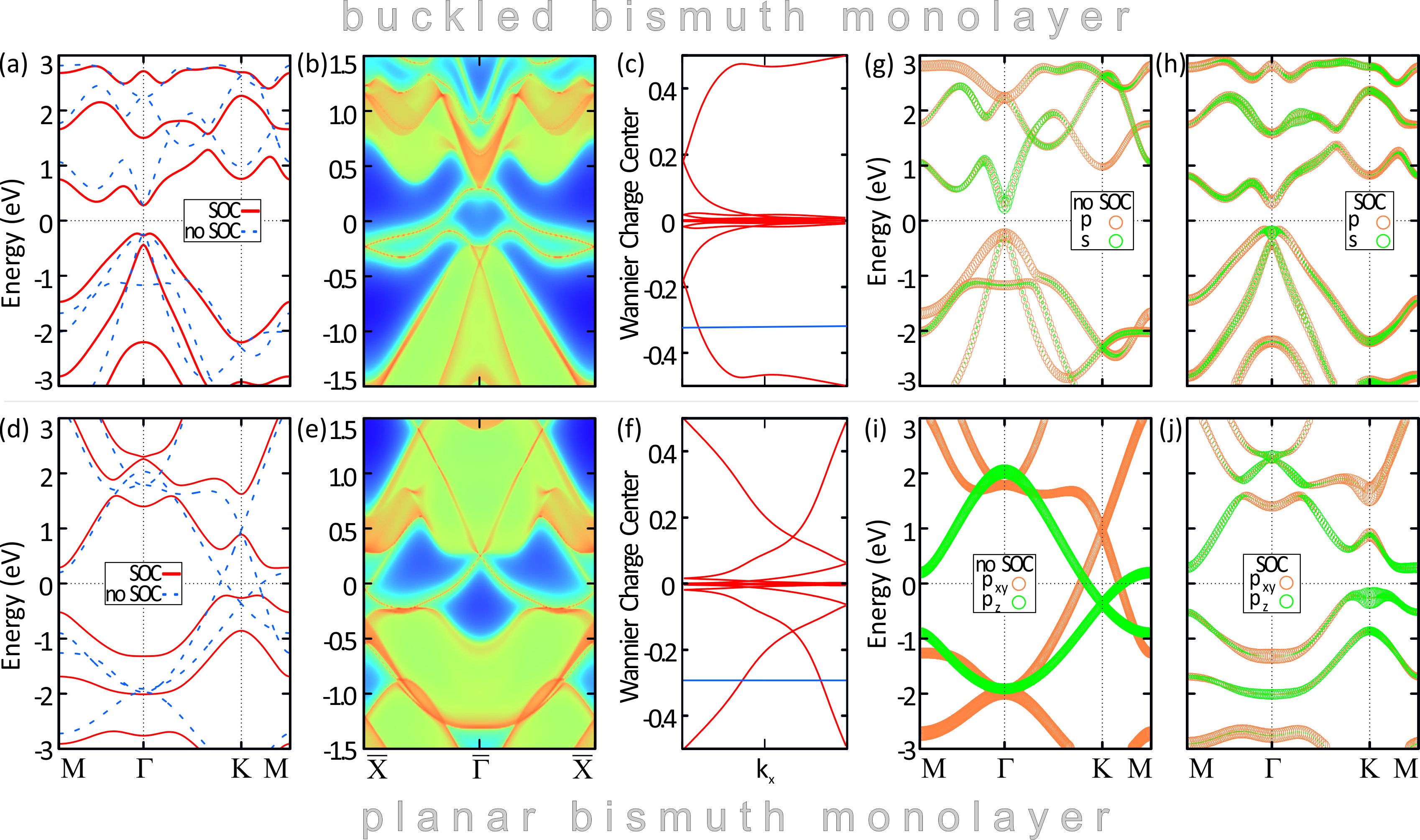}
\caption{(a) The electronic structure, (b)  zigzag edge states and (c)  Wilson loop analysis of the freestanding bismuth in the buckled geometry. (d), (e) and (f) correspond to the same quantities in (a-c) for the planar geometry. The solid red and blue dashed line correspond to calculations with/without SOC in (a),(d). The solid blue line in (c)/(f)crosses the Wannier charge center odd/even times, corresponding to the topologically trivial/nontrivial case for buckled/planar bismuth, respectively.   (g) and (h) display the $s$ (green) and $p$ (orange) orbital projections for the buckled bismuth monolayer without and with SOC; (i) and (j) are same as (g) and (h) but for the planar bismuth. The orange (green) color corresponds to the $p_{xy}$ ($p_{z}$) orbitals.}
\label{monolayer-bi}
\end{figure*}

\section{Electronic structure in (Bi, Sb, As)/SiC}
\label{Sec:ElectronicStructure}

In this section, on the basis of a priori DFT calculations, we begin to elaborate on the step-by-step analysis of the low-energy electronic structure of a monolayer made of group-V elements Sb, As, and Bi deposited on a honeycomb substrate, in this case SiC. 

As the first pivotal insight into the effective electronic model as it appears in $H_{0}^{\sigma\sigma}$ of Eq.~(1), \textit{the substrate imposes an orbital filtering mechanism}: it strongly binds to the $p_z$ orbital of the group-V monolayer, and as such removes it from low energies below or above the Fermi level. Furthermore, Dirac cones at the K (K') point form the low-energy electronic dispersion, and within a good, quantifiable approximation solely are of $p_x$ and $p_y$ orbital nature. 
We first present in Sec.~\ref{freestanding} the band character and topology analysis of the freestanding monolayer bismuth to highlight the role of the substrate.
Next, the DFT electronic structure results of the combined (Bi, Sb, As)/SiC layer plus substrate systems are presented in Sec.~\ref{material_aspect}.
This allows us to construct a universal low-energy theory using a symmetry analysis of the monolayer-substrate heterostructure, followed by an electronic structure analysis that initially does not include SOC. The inclusion of the SOC will be followed up on in Sec.~\ref{Sec:LowEnergyModel}.
Unless explicitly stated otherwise, our first-principle band-structure calculations were carried out within DFT as implemented in the Vienna Ab-initio Simulation Package~\cite{kresse1996} (VASP). More details of our calculations can be found in Appendix~\ref{appendix:DFTdetails}.
\subsection{Freestanding bismuth monolayer}\label{freestanding}

To better understand the pivotal role played by the substrate for the emergence of the large gap QSH effect in the (Bi,Sb,As)/SiC combined systems, we show the bulk electronic structure, the zigzag edge states and the Wilson loop (calculation of the $Z_2$ invariant)  analysis for the prototype example of a freestanding bismuth in Fig.~\ref{monolayer-bi}(a-c) for the buckled  configuration and in Fig.~\ref{monolayer-bi}(d-f) for the planar configuration, correspondingly. 
These two different configurations are both free of a substrate and preserve inversion symmetry, which results in the Kramers' degeneracy at every k-point. 
Before going into more detailed discussions, we first present our conclusions and emphasize that $p_{z}$ bands participate in the low-energy sectors in both monolayer cases. First, a substrate like SiC applies a tensile strain to the bismuth monolayer, which induces a planar geometry (Fig.~2) and changes the fundamental gap from $\Gamma$ to K. Second, the topology of both monolayer systems are derived from band inversions. There is an $s$-$p$ band inversion at $\Gamma$ in the buckled geometry (Fig.~\ref{monolayer-bi}(g) and (h)), which leads to the buckled monolayer bismuth being a QSH insulator with $Z_{2}=1$. 
While planar bismuth is not a QSH insulator protected regularly by time-reversal symmetry, still an inversion between $p_{xy}$ and $p_{z}$ band complexes still exists (Fig.~\ref{monolayer-bi}(i) and (j)). 
The Wilson loop analysis, however, displays a trivial structure in this case.

In both cases, a gap of 0.5 eV is observed in the bulk electronic structure in the presence of the large SOC of Bi (red lines in Fig.~\ref{monolayer-bi}(a) and (d)). 
The fundamental difference of them lies in the gap location. 
Buckled bismuth has its gap at $\Gamma$, with an inverted band order between Bi-$s$ and Bi-$p$ bands induced by the SOC (see the blue dashed line in Fig.~\ref{monolayer-bi}(a) for the case of no SOC). Here, the topological edge states and a nonzero topological invariant $Z_{2}=1$ are easily derived (see Fig.~\ref{monolayer-bi}(b) and (c) as well as  Wilson analysis in App. C). This buckled system is  a free-standing material which, according to the DFT (calculations presented in App. A) is a stable geometry . However, so far it has not been realized ecperimentally. More generally, it is clear that, for all practical applications,
a free-standing film must eventually be placed or grown on a substrate.
  
Applying external tensile strain to the buckled bismuth significantly modifies the bulk electronic structure and eventually renders the material planar. 
Without SOC, the two linear band crossings at K appear at above and below the Fermi level. They carry distinct orbital characters, with the one above the Fermi level consisting of Bi-$p_{x}$ and $p_{y}$ orbitals while the one below the Fermi level mainly consisting of Bi-$p_{z}$ character (Fig.~\ref{monolayer-bi}(i)). 
$p_{xy}$ and $p_{z}$ bands switch order at K and are further gapped  by introducing SOC (Fig.~\ref{monolayer-bi}(i,j)).
A direct inspection of the Wilson loop (Fig.~\ref{monolayer-bi}(f)) concludes that this system still is a trivial insulator ($Z_{2}=0$) consistent with a previous analysis (see Ref.~\onlinecite{bansil} and references therein).  
In addition to the Wilson loop analysis, one can analyze the structure of edge states for both types of monolayers between $\overline{\Gamma}$  and  $\overline{X}$ points in the Brillouin zone (compare Fig.~\ref{monolayer-bi}(b) with Fig.~\ref{monolayer-bi}(e)).
There is an odd number of crossings for the buckled bismuth monolayer (i.e. topological situation) and an even number (two) crossings for the planar configuration (blue line in Fig.~\ref{monolayer-bi}(c) and (f)). 

Further comparison of Fig.~\ref{monolayer-bi}(a) and (d) with the electronic structure of Bi/SiC shown in our previous work~\cite{Reis2017:S} as well as in Fig.~\ref{Fig:Bi_band} reveals that the presence of SiC substrate does a two-step job: (i) it stretches the bismuth monolayer and makes it completely flat, which changes the topological gap from $\Gamma$ to K, and separates the $p_{xy}$ from the $p_{z}$ bands (see Fig.~\ref{monolayer-bi}(a) and (d)), and (ii) it removes the $p_{z}$ bands completely from the low-energy region close to the  Fermi level and {\it transforms the topology of the system from a band-inversion type to graphene type, while still preserving the large topological gap.} Thus, as in the KM model, the hexagonal lattice structure conspires to form a topological gap, albeit with this gap being determined by the large on-site SOC in our case.

\begin{figure}[htbp]
\centering
\includegraphics[width=\linewidth]{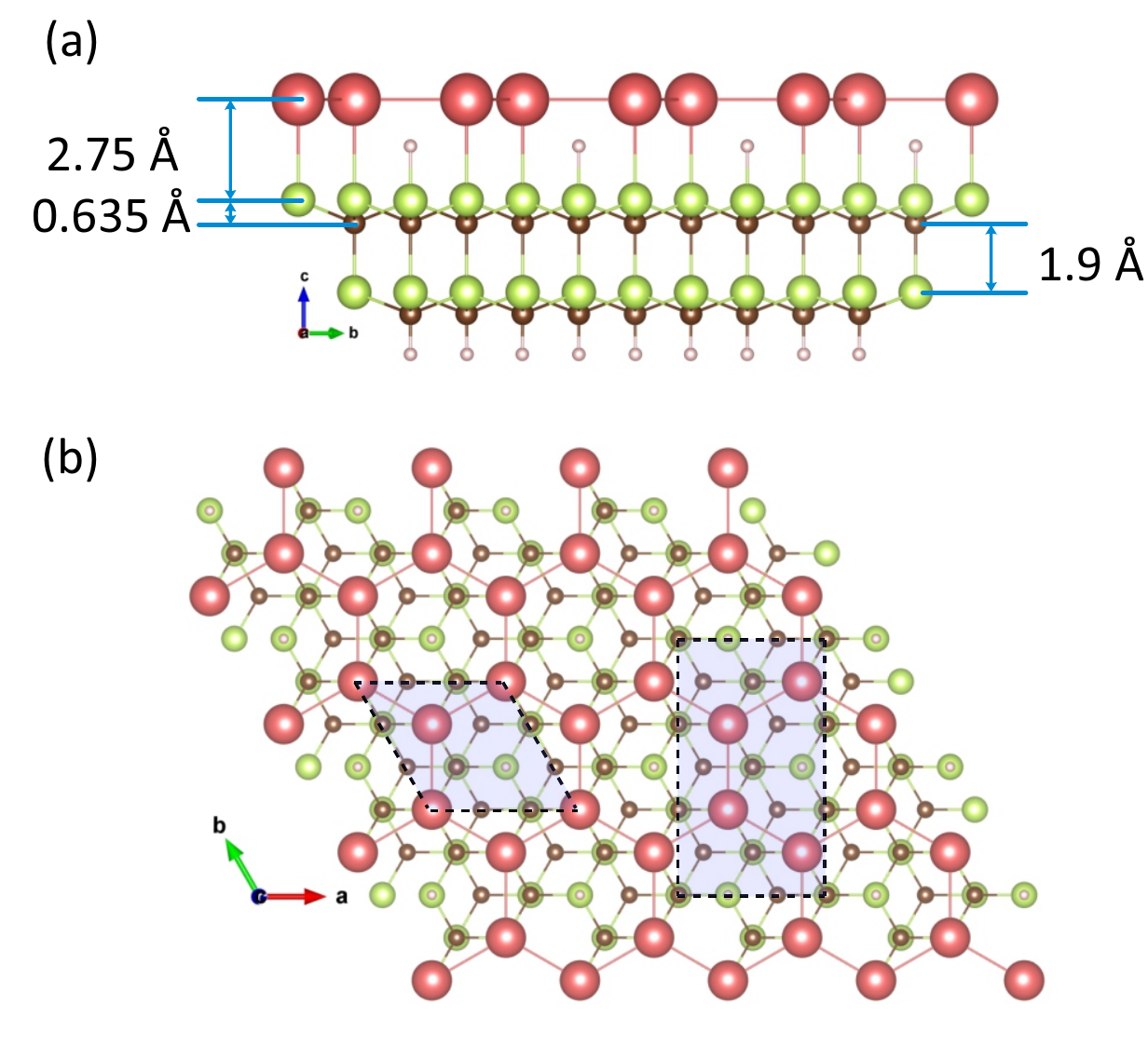}%
\caption{(a) Side and (b) top views of the Bi/SiC(0001)-$\sqrt{3}$ monolayer structure. The diamond light-blue cell is the unit-cell, whereas the rectangle cell is a conventional cell which is arranged in zigzag and armchair shapes along x- and y- directions, respectively.}
\label{Fig:lattice_structure}
\end{figure}

\subsection{Heterostructure and material properties}\label{material_aspect}
\subsubsection{The bismuthene/SiC system}
\begin{figure}[h]
\centering
\includegraphics[width=0.8\linewidth]{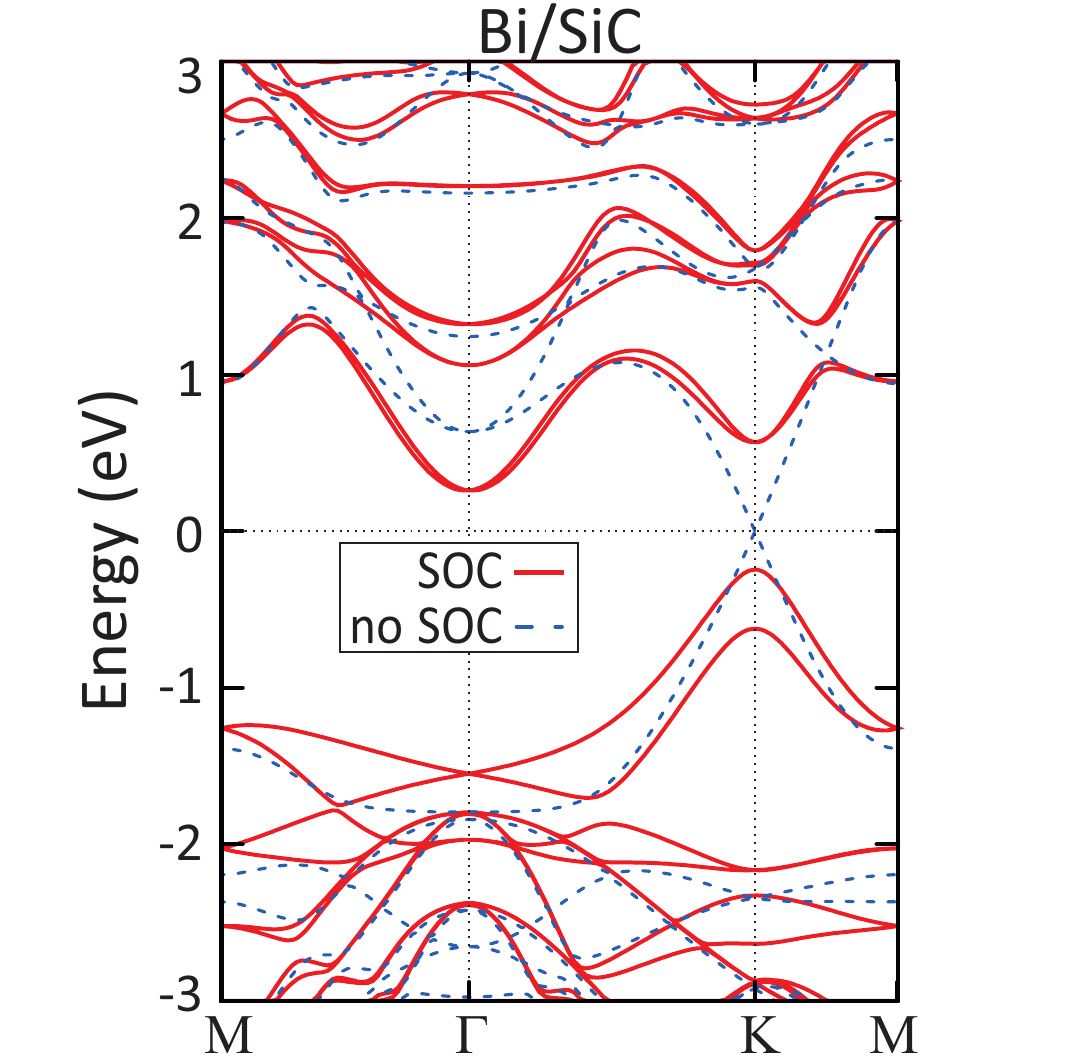}
\caption{Electronic structure of the Bi/SiC(0001) with the blue-dashed line and the-red solid line corresponding to the cases of without and with SOC.}
\label{Fig:Bi_band}
\end{figure}

Fig.~\ref{Fig:lattice_structure} presents our concept for bismuthene on top of the SiC substrate, which is described in detail in our earlier work.\cite{Reis2017:S} In the upper part of Fig.~\ref{Fig:lattice_structure} a side view is displayed, where one can see an important structural influence of the substrate: it induces a fully planar configuration of the honeycomb rings.
This planar collapse (lattice constant 5.35 \AA) is due to a sizable tensile strain of 18\%, compared to buckled (111) layers. The lower part of Fig.~\ref{Fig:lattice_structure} contains the top view with further details of the atomic composition of our $\sqrt{3}\times\sqrt{3}$ R $30^\circ$ Bi/SiC(0001) system, which we amply be employed in the subsequent discussion.
Fig.~\ref{Fig:Bi_band} summarizes the a-priori electronic structure results, based on DFT calculations, for this Bi/SiC material realization. The blue-dashed line in this figure shows the band structure without spin-orbit coupling, whereas for the red-solid line the SOC is included. 
A bulk Dirac-band crossing appears at the K-point, when the spin-orbit coupling is not present, whereas this crossing is fully gapped out in the calculations including SOC, where a topological gap appears. Among all so far proposed topological insulators, we note that the bismuthene/SiC combination displays the largest bulk energy gap, so desirable in view of possible applications. This is shown in both the (qualitatively rather similar GGA and HSE) exchange-correlation functional implementations, with a gap of size 0.816 eV (0.956 eV) opened at the K point in GGA (HSE), respectively. The fundamental ({\it i.e.} indirect) gap is smaller, 0.506 eV (0.668 eV) in the GGA (HSE) calculations (App.~\ref{appendix:DFTdetails}).

\begin{figure}[htbp]
\centering
\includegraphics[width=0.8\linewidth]{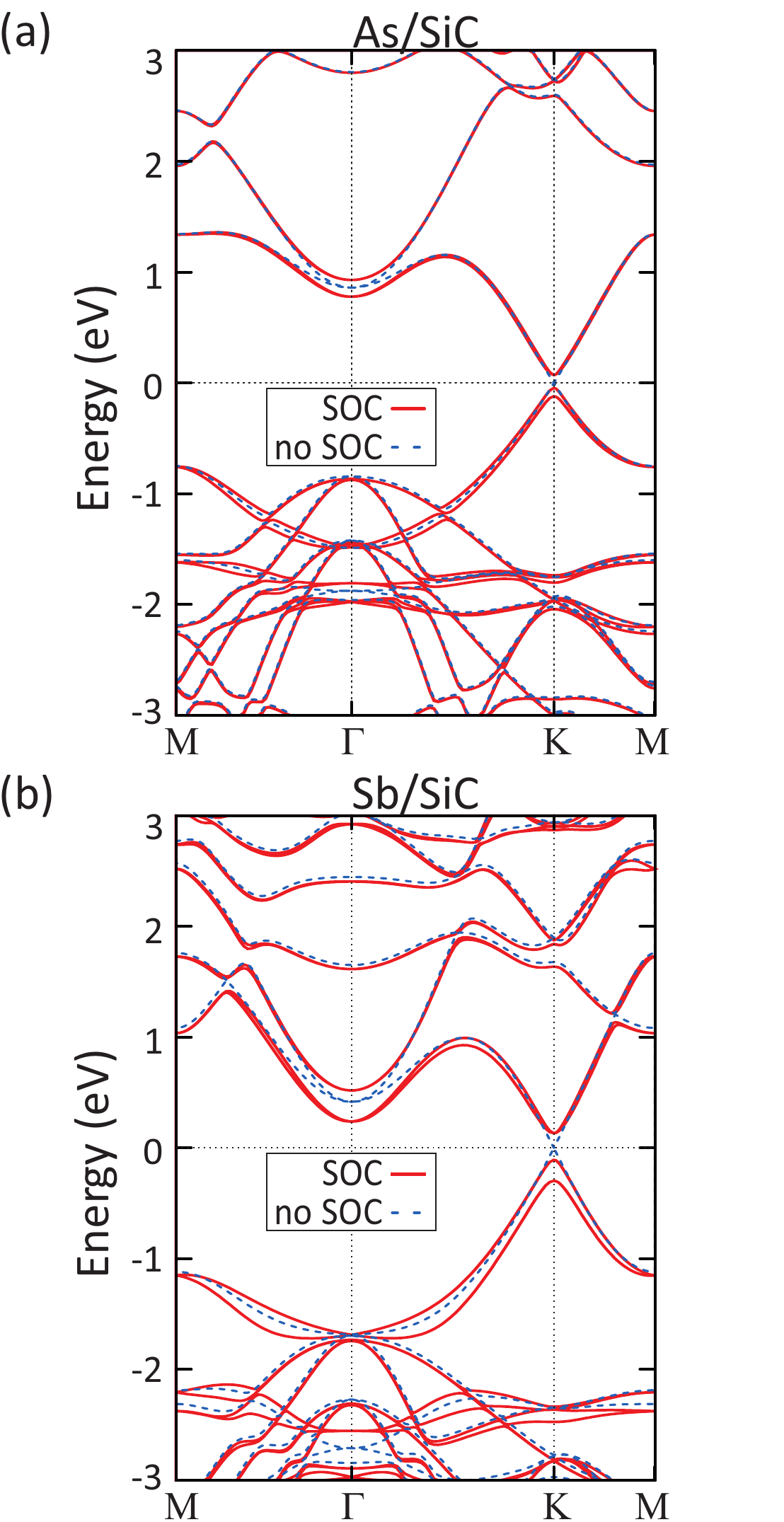}
\caption{Electronic structures of arsenic and antimony on SiC(0001). For each figure, the left and right plots correspond to the calculations with and without spin-orbit couplings.}
\label{SbAs_band}
\end{figure}

\subsubsection{Comparison with (Sb, As)/SiC}

In the Supplement of our recent work\cite{Reis2017:S}, we have provided for the Bi/SiC (0001) material realization a first derivation of the effective low-energy model starting from an a-priori DFT band structure calculation. Here, in Secs.~\ref{Sec:ElectronicStructure} and~\ref{Sec:LowEnergyModel}, the emphasis is on generalizing this low-energy description to other heavy-atom monolayer/substrate combinations, thereby setting up the paradigm.
Figs.~\ref{SbAs_band}(a,b) predict, on the basis of DFT calculations (plus an explicit evaluation of the topological constant $Z_2$ in Appendix~\ref{appendix:TopInv}), that a similarly large topological gap occurs in the Sb/SiC and As/SiC systems as in the Bi/SiC combination (Sb/SiC $\sim0.3$ eV, As/SiC $\sim0.2$ eV). We also detect the Rashba band splitting in the valence bands, acting as in Bi/SiC as a possible consistency check ({\it e.g.} in ARPES experiments) for the topological properties induced by the SOC.
As for Bi/SiC, the monolayers of (As, Sb) are taken to be flat, see Appendix~\ref{appendix:DFTdetails}. The free-standing (As,Sb)-layers, however, have a lattice constant (3.6 {\AA} for As, 4.1 {\AA} for Sb) significantly smaller than that of SiC and, therefore, the tensile strain is rather large, much larger than in Bi/SiC\cite{Reis2017:S}. 

This assessment is based on comparing the total energy curves via lattice constants for a hypothetical free-standing material, {\it i.e.} (i) the buckled monolayer and (ii) the planar monolayer (App.~\ref{appendix:DFTdetails}). While the former was found in Bi/SiC to have its minimum at 4.33 {\AA}, the hypothetical freestanding planar layer has its minimum at 5.27 \AA.\cite{Huang2013:PRB} Our "real-world" system of planar bismuthene on a SiC substrate has a lattice constant of 5.35 {\AA}.\cite{Reis2017:S} Thus, the actual lattice constant is very close to the energy minimum for the freestanding planar case. Thus, our Bi/SiC combination is not subject to a strain issue, which is an important finding in our earlier work.
On the other hand, in the (Sb, As) systems, the lattice constants (Appendix~\ref{appendix:DFTdetails}) need to be significantly enlarged, {\it i.e.} from $a=3.6$ {\AA} to $a=4.37$ {\AA} in As, and from $a=4.1$ {\AA} to $a=5.1$ {\AA} in Sb, to achieve the planar configuration. While alternative substrates would suggest themselves for further investigation, in this work we confine ourselves to the (As,Sb)/SiC systems, assuming planar layers.
Another point to note is that the SiC substrate gap is 3.2 eV and, thus, large enough to accommodate, at least in principle, the large topological gaps ({\it i.e.} in bismuthene (0.8 eV), and in our above (Sb/As) systems with gaps of order (0.3, 0.2 eV)), where $E_F$ lies within this topological gap.

In summary of this (Sb, As)/SiC comparison with the Bi/SiC system, we can already detect a unifying aspect: it is the reduced atomic SOC strength of Sb and As compared to Bi, which is responsible for the only qualitative difference of the low-energy band structure and, in particular, the gap results, {\it i.e.} the size of the topological gap and the Rashba valence-band splitting shrink. 
This is in accordance with our paradigm, in that it is the layer-substrate bonding, which allows for the large on-site SOC to directly come into play, generating gaps of the order of several hundreds meV.  This central aspect will be considered next. 

\subsubsection{The Role of the Substrate}

\begin{figure}[htbp]
\centering
\includegraphics[width=\linewidth]{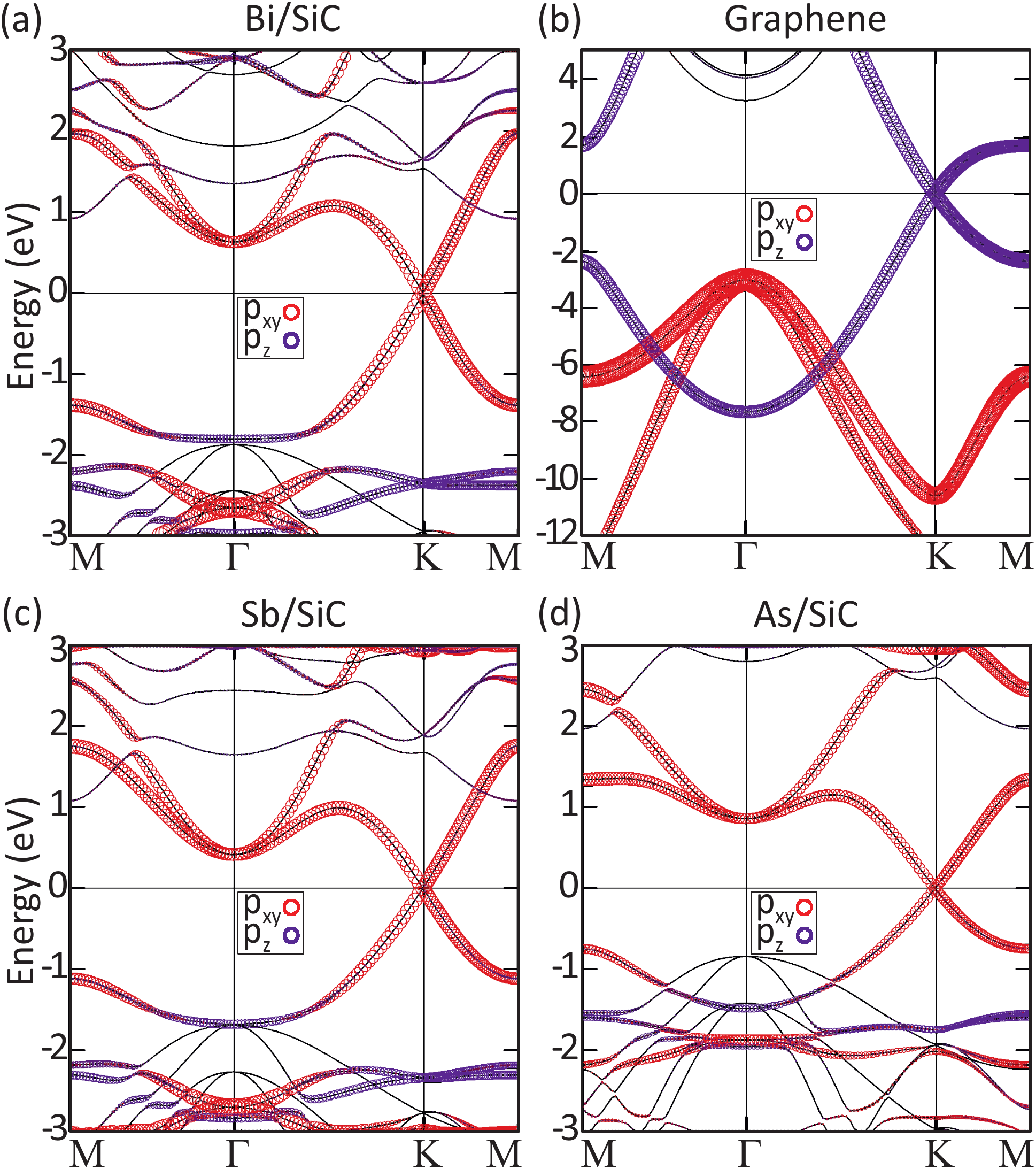}
\caption{The projection of the electronic structure to the three $p$ orbitals of (a) Bi in Bi/SiC(0001); (b) C in graphene; (c) Sb in Sb/SiC(0001) and (d) As in As/SiC(0001). The zero energy level corresponds to the Fermi energy.}
\label{band_decomposition}
\end{figure}
Let us start with Fig.~\ref{band_decomposition}, which in (a) displays the orbital-resolved electronic structure of the DFT calculation for our prototype example Bi/SiC (0001), first without inclusion of SOC. In Fig.~\ref{band_decomposition}, the circle size is proportional to the relative weight of the orbital. In this monolayer/substrate system, the linear bands which cross at the K-point, consist mainly of $p_x$ and $p_y$ orbitals.
In contrast, in Fig.~\ref{band_decomposition}(b), which plots the bands and the orbital projections for graphene, the low-energy physics is due to just one orbital, {\it i.e.} the $p_z$-orbital.
Obviously, the electronic structure of the quasi-2D heavy-atom system, which comprises $s$- and $p$-orbitals of {\it e.g.} the Bi-atoms, is substantially modified by the presence of the substrate: as a free standing layer, Bi-atoms would form $sp^2$ bands, leading to the $\sigma$-bands of bismuthene, while the ``dangling" $p_z$-orbitals point out of the plane and give rise to the $\pi$-bands. In this case (see Fig.~\ref{band_decomposition}(b)), the low-energy states around $E_\text{Fermi}$ have $p_z$ orbital character.

\begin{figure}[htbp]
\centering
\includegraphics[width=\linewidth]{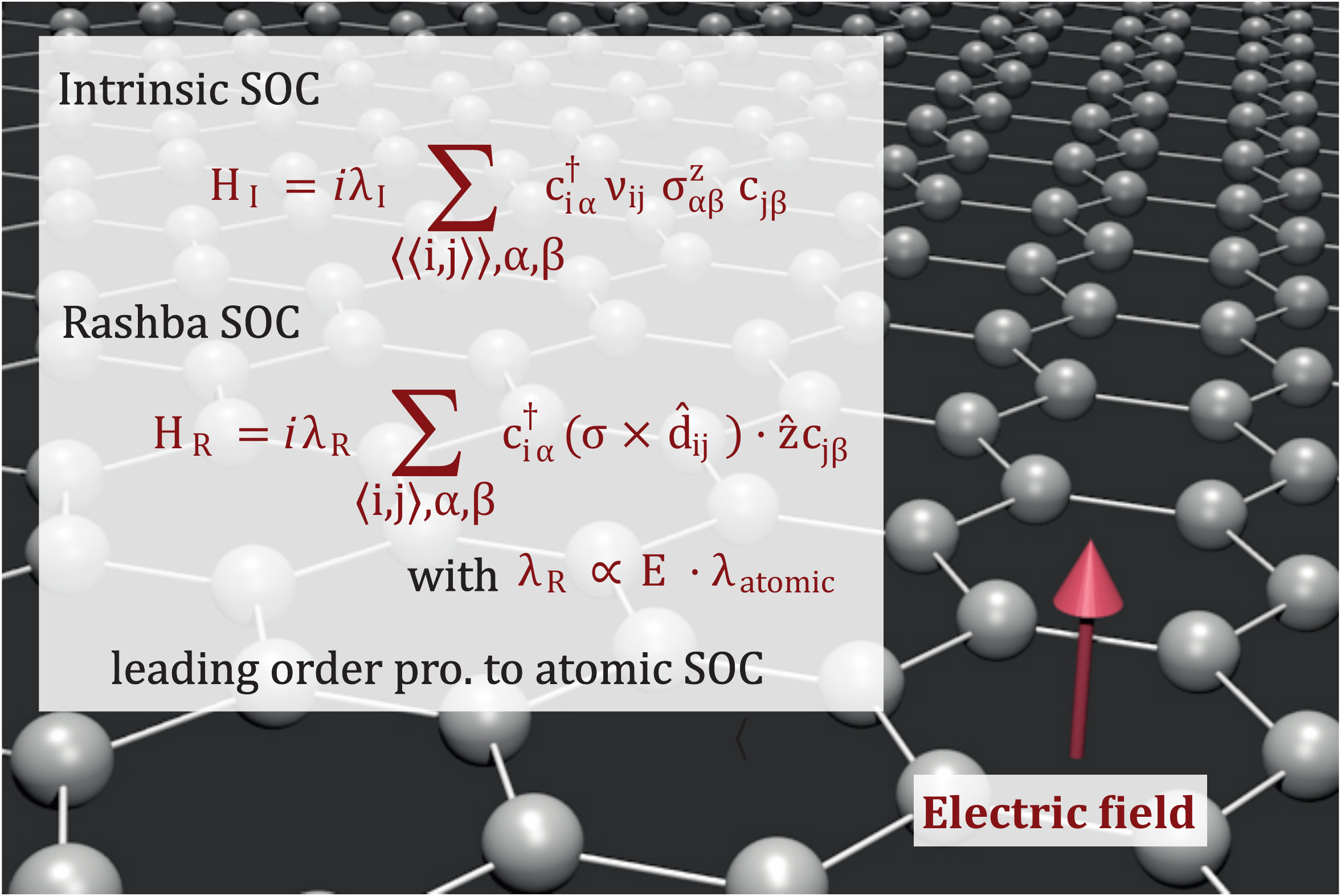}
\caption{``Gedanken" experiment to probe the Rashba SOC.}
\label{Gedanken}
\end{figure}

As known, from the Kane-Mele work on graphene,\cite{kane-05prl226801} this single $p_z$-band in the honeycomb lattice gives rise to a tiny band structure SOC at the level of higher-order perturbation theory (Fig.~\ref{Gedanken}). However, not surprisingly, it is precisely this "dangling-band" $p_z$-orbital band, which is most substantially affected by the presence of the substrate: its bonding to the honeycomb layer acts like an electric gate field $E$ imposed on the Bi(As,Sb)-orbital manifold and shifts the $p_z$-states away from the low-energy sector of the combined layer/substrate system.
This is summarized in a type of Gedankenexperiment in Fig.~\ref{Gedanken}, where a perpendicular electric field is assumed to act on the honeycomb layer (see also Ref.~\onlinecite{PhysRevB.74.165310}): the general idea here is, that the effect of the honeycomb-layer bonding to the substrate can be effectively absorbed into an electric field $\vec{E}$, applied perpendicular to the layer. The explicit construction is presented in App.~\ref{appendix:DFTdetails}, where the coupling strength $\lambda_{E}$ along the $z$-direction can directly be determined from a tight-binding fit to the DFT results. Here, it is used to develop a simple-as-possible insight into the role of the substrate.  

For the Kane-Mele scenario in Fig.~\ref{Gedanken}, with only the $\pi$-band, the matrix element of the intrinsic SOC $H_I$ obviously vanishes at the same site, {\it i.e.} for $i=j$, if only the $p_z$-orbital is involved [the well-known forms of $H_I$ for the intrinsic and $H_R$ for the Rashba SOC, are defined in detail in Eqs.~(7,8) in the App.~\ref{appendix:LowEnergyModel}]. The substrate, or the electric field $E$, then projects the $\pi$-band to high energy. This is visible in Fig.~\ref{band_decomposition}(a) for Bi/SiC and can be shown similarly for the new As/SiC system, where the dominant low-energy states are now due to As-$p_x$ and As-$p_y$ orbitals. The $\bm{L}\cdot\bm{S}$ SOC, where $\bm{L}$ and $\bm{S}$ denote the orbital and spin angular momenta, then gives rise to the large atomic on-site SOC ({\it i.e.} $\lambda_\text{SOC}H_\text{SOC}^{\sigma\sigma}$ in Eq.~(\ref{EffLowEnergyHamiltonianGeneral})), due to the $L_z\sigma_z$-term, which connects $p_x$ with $p_y$ orbitals (Sec.~\ref{Sec:LowEnergyModel} below).
This crucial effect of the substrate establishes the "high-temperature" QSH paradigm, in that the honeycomb layer (Bi,As,Sb etc.) substrate ({\it e.g.} SiC) combination now displays a systematic scaling in its band gap with the large magnitude of the atomic SOC of the heavy elements.

The phenomenology of modifying the low-energy behavior to the dominance of $p_x$- and $p_y$-orbitals is often termed as ``orbital filtering", or, ``orbital engineering".
It has previously been used in a variety of systems such as cold-atom lattices, where a laser beam acts as an external field and shifts the $p_z$-orbital band to high energy \cite{ZhangGF2014}. In this way, as discussed in our ``Gedanken" experiment, the Kane-Mele situation, with just the $p_z$ orbital band creating the tiny topological gap can be avoided. A somewhat similar strategy is considered in a large variety of recent DFT electronic structure calculations of functionalized heavy-metal atomic layers,\cite{bansil,PhysRevB.90.085431} {\it i.e.} placing hydrogen atoms on one side of the planar honeycombs, or heavy-metal atoms (such as Bi) on top of halogen-covered Si-surfaces.\cite{zhou}
A nice review of these and related DFT calculations is contained in a recent work by C.-H. Hsu et al.,\cite{bansil} where already DFT calculations of Bi/SiC and Sb/SiC were found to support a large non-trivial band gap. 
These important calculations revealed the presence of edge states; however, they did not derive the low-energy effective Hamiltonian, which is necessary to resolve the crucial role of the on-site, {\it i.e.} intrinsic SOC and the Rashba term, as explained in Sec.~\ref{Sec:LowEnergyModel} and, in more detail, in Appendix~\ref{appendix:LowEnergyModel}.

For the latter term, we consider explicitly that the substrate breaks inversion symmetry and, thereby, creates a Rashba SOC (see the term $\lambda_\text{R}H_\text{R}^{\sigma\sigma}$ in Eq.~(\ref{EffLowEnergyHamiltonianGeneral})), in addition to the above ``intrinsic" SOC. From our earlier ``Gedanken" Experiment in Fig.~\ref{Gedanken}, we know that the Rashba term scales with $\lambda_\text{R}\sim E\lambda_\text{atomic}$. Thus, depending on the bonding strength ($E$) between layer and substrate, we can expect a large (in the Bi/SiC case $\sim0.4$ eV) Rashba splitting, easily detectable via ARPES in the valence bands. This splitting varies from one layer/substrate system to the next, as can be seen from Figs. 4(a,b) for the Sb/SiC and As/SiC systems. Let us now go, step-by-step, through these findings in the context of our low-energy description.

\subsection{Low-Energy Model without Spin-Orbit Coupling}
In the Slater-Koster treatment without SOC, we start with only $p_x$- and $p_y$-orbitals creating the low-energy description of the $\sigma$-bands, {\it i.e.} with the four basis functions for spin up ($\uparrow$)
\begin{equation}\label{basisfunction}
|p_{x\uparrow}^A\rangle,|p_{y\uparrow}^A\rangle,|p_{x\uparrow}^B\rangle,|p_{y\uparrow}^B\rangle\;,
\end{equation}
and the same four basis functions but for spin down ($\downarrow$).

A and B denote the two inequivalent sites in the honeycomb unit cell. The direct and reciprocal lattice vectors, as well as the high-symmetry points in the BZ of the honeycomb lattice, {\it i.e.} $\Gamma$, $M$, K and K' take the usual values (see  Appendix~\ref{appendix:LowEnergyModel}). The Hamiltonian ($4\times4$) matrices in the two sectors are equivalent and can straightforwardly be expressed in terms of on-site ``$AA$" and ``$BB$" as well as nearest-neighbor ``$AB$" Slater-Koster (SK) integrals, {\it i.e.} Eq.~(\ref{H1}).
\begin{equation}\label{H_uu}
H_{\uparrow\uparrow}= H_{\downarrow\downarrow}=
\begin{pmatrix}
0					 & 	0 					& h_{xx}^{AB} & h_{xy}^{AB}\cr
0					 & 	0 					& h_{yx}^{AB} & h_{yy}^{AB} \cr
\dagger 		& \dagger 			& 0 					& 0 \cr
\dagger 		& \dagger 			& 0 					& 0
\end{pmatrix}\;\text{,}
\end{equation}
where $\dagger$ denotes the complex conjugate of the matrix elements of the terms between sites $A$ and $B$ shown in the upper matrix. The eigenenergies at the K-point are then:
\begin{equation}
E=0,0,\pm\frac{3}{2}\left(V^1_{pp\pi}-V^1_{pp\sigma}\right)
\end{equation}
Here (see also Eq.~(\ref{SK_integral}), the $AA$, $BB$ and $AB$ SK-integrals above have further been decomposed into onsite integrals, {\it i.e.} $V^0_{pp\sigma}=V^0_{pp\pi}=0$ with our choice of zero energy ($E_{F}=0$) at the Dirac crossing and nearest-neighbor $V^1_{pp\sigma}$ and $V^1_{pp\pi}$ overlap integrals. Thus, two of the states are degenerate at the K-point (Fig.~\ref{Fig:Bi_band} and Fig.~\ref{effective_model}, left-hand panel).

\section{Relativistic low-energy Model}
\label{Sec:LowEnergyModel}

Building upon $H_{0}^{\sigma\sigma}$ from Sec.~\ref{Sec:ElectronicStructure}, we include SOC into the effective model by performing a relativistic DFT analysis. The $p_x, p_y$ two-orbital basis at low energies triggers a Dirac gap opening at the K (K') point implied by intrinsic atomic SOC $H_{\text{SOC}}^{\sigma\sigma}$. Furthermore, by including matrix elements between the $\sigma$-bond and $\pi$-bond sector, we find that the substrate induces a Rashba SOC term $H_{\text{R}}^{\sigma\sigma}$, which splits the previously degenerate nearby valence band at K (K'), but not the conduction band. Together with Sec.~\ref{Sec:ElectronicStructure}, this illustrates how fundamentally the substrate modifies the low-energy electronic structure of the monolayer. We provide a synopsis of the (As,Sb,Bi)/SiC heterostructure compounds.

\subsection{Intrinsic and Rashba Spin-Orbit Coupling}
\begin{figure}[htbp]
\centering
\includegraphics[width=\linewidth]{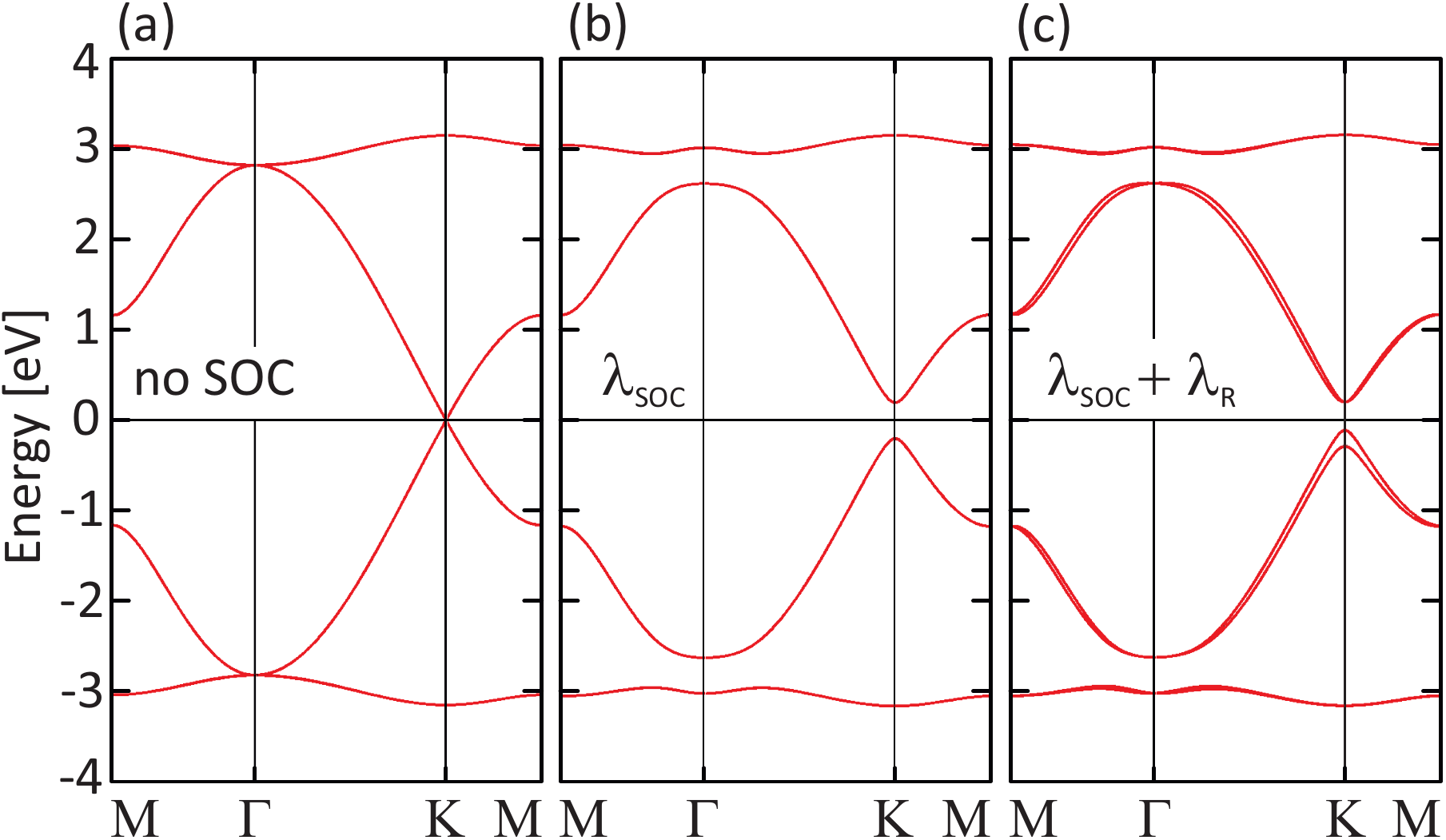}
\caption{The electronic structure of the low-energy effective model for Sb/SiC $\sigma$-bands. a) $\lambda_{so}=\lambda_{R}=0$, b) $\lambda_{so}=0.2, \lambda_{R}=0$, c) $\lambda_{so}=0.2, \lambda_{R}=0.015$.}
\label{effective_model}
\end{figure}

As discussed before, the ``orbital filtering" due to the presence of the substrate allows now in the $\sigma$-orbital sector (as defined in Eq.~(\ref{basisfunction}) in Sec. II B), for an on-site ({\it i.e.} atomic) intrinsic SOC, arising from the $L_z\sigma_z$-term in the atomic spin-orbit coupling,
\begin{equation}
\lambda_\text{SOC}\bm{L}\cdot\bm{S}.
\end{equation}
Taking this term into account, which mixes $|p_x\rangle$- and $|p_y\rangle$-basis functions, {\it i.e.}
\begin{equation}\label{intrinsicLS}
\langle p_y|\bm{L}\cdot\bm{S}|p_x\rangle=\i\sigma_z,\quad\langle p_x|\bm{L}\cdot\bm{S}|p_y\rangle=-\i\sigma_z,
\end{equation}
we straightforwardly obtain the Hamiltonian matrix for the $\sigma$-bands, as detailed in Eq.~(\ref{Hss}). In the presence of this SOC, the (4$\times$4) matrix description of Eq.~(\ref{H_uu}) is augmented to an (8$\times$8) Hamiltonian matrix. Because the $L_z\sigma_z$-term does not mix the different spin sectors, the band structure still comprises only four bands, two valence and two conduction bands, separated by the large topological gap at the K-point (see the middle panel in Fig.~\ref{effective_model}).

This physics is also captured in the orbital decomposition of the electronic structure of our V/SiC (with V = Bi, Sb, As) system: from Fig.~\ref{band_decomposition}(a) we note, that the "orbital filtering" is not uniform in the BZ. While the band crossing around Dirac bands is indeed dominated by $p_x$- and $p_y$-orbital weights, closer inspection reveals that one of the two topologically relevant bands, {\it i.e.} the top valence band still carries a large weight of Bi(As,Sb)-$p_z$ character around the $\Gamma$-point. We will see now that both this ``high-energy" $p_z$-contribution and the Bi(As,Sb)-s orbital weight are crucial for setting up, finally, the Rashba SOC.
As a consequence, we have to start from an orbital basis which, in addition to the ``low-energy" orbitals $p_x$ and $p_y$ relevant at the Dirac K-point, has to include the Bi (As,Sb)-$s$ and Bi (As,Sb)-$p_z$ orbitals forming the $\pi$-bands.

Details of the downfolding, including these latter orbitals, can be found in App.~\ref{appendix:LowEnergyModel} (Eqs.~(\ref{Hss}) to~(\ref{Hpp})). Here, we shortly summarize the results: in this complete basis the ($L_x\sigma_x +L_y\sigma_y$) SOC becomes relevant which couples the $p_x$-, $p_y$-orbitals with the "high-energy" $p_z$-orbital. In contrast to the $L_z\sigma_z$ term, which connects $p_x$ with $p_y$-orbitals (see Eq.~(\ref{intrinsicLS})) and which gives rise to the large on-site SOC ($\lambda_\text{SOC}$), the ($L_x\sigma_x +L_y\sigma_y)$ term mixes the different spin sectors in $H^{\sigma\pi}$ and $H^{\pi\sigma}$, see Eqs.~(\ref{Hmix})-(\ref{Hsp}).

The full Hamiltonian matrix (16$\times$16) is spanned by both Bi (As,Sb)-$s$ orbitals and the Bi (As,Sb)-$p$ orbitals, including their spin and sublattice degrees of freedom. In the final step, to obtain an effective (8$\times$8) low-energy $\sigma$-band model, second order perturbation theory is applied which includes the $\pi\sigma$ hybridization within the $\sigma$-band subspace, {\it i.e.}, see Eq.~(\ref{Hsp}).
\begin{equation}
H_{\text{eff}}^{\sigma\sigma}\approx H^{\sigma\sigma}-H^{\sigma\pi}\cdot\left(H^{\pi\pi}\right)^{-1}\cdot H^{\pi\sigma}
\end{equation}
In summary, we arrive at the effective low-energy model as in Eq.~(\ref{EffLowEnergyHamiltonianGeneral}) for the combined hexagonal heavy-atom layer, substrate system.

The intrinsic SOC-term ($\sim\lambda_\text{SOC}$) and the Rashba term ($\sim\lambda_\text{R}$) are explicitly extracted from the above results for $H_\text{eff}^{\sigma\sigma}$ in Eq.~(\ref{effective_H}), where $H_\text{eff}^{\sigma\sigma}$ is defined as:
\begin{equation}
H_{\text{eff}}^{\sigma \sigma} =
\begin{pmatrix}
H_{\uparrow\uparrow}^{\sigma \sigma} & H_{\uparrow\downarrow}^{\sigma \sigma} \cr
H_{\downarrow\uparrow}^{\sigma \sigma} & H_{\downarrow\downarrow}^{\sigma \sigma}
\end{pmatrix}\;,
\end{equation}

The first two contributions in Eq.~(\ref{EffLowEnergyHamiltonianGeneral}) then derive from the term:
\begin{equation}
H_{\uparrow\uparrow/\downarrow \downarrow}^{\sigma \sigma}=H_{0,\uparrow\uparrow/\downarrow \downarrow}^{\sigma \sigma}\pm \lambda_{\text{SOC}} \begin{pmatrix}
0 & -i & 0 & 0\cr
i & 0 & 0 & 0 \cr
0& 0 & 0 & -i\cr
0 & 0 & i & 0
\end{pmatrix},
\end{equation}
where the matrix in the second term stands for the intrinsic SOC in the corresponding spin sector, {\it i.e.} $H_\text{SOC}^{\sigma \sigma}$. The Rashba effect, on the other hand, is contained in the matrix
\begin{equation}
H_{\uparrow\downarrow}^{\sigma \sigma} = (H_{\downarrow\uparrow}^{\sigma \sigma})^{\dagger}=
\lambda_{\text{R}}\begin{pmatrix}
0 & 0 & a & b\cr
0 & 0 & b & c \cr
d & e & 0 & 0 \cr
e & f & 0 & 0
\end{pmatrix}\; ,
\end{equation}
where the elements $a$ to $f$ are defined in Eq.~(\ref{Coefficient}).

As can be seen in Fig.~\ref{effective_model}, this low-energy model correctly reproduces the ARPES band structure. In particular, it allows, via the Rashba splitting in the valence band, a crucial consistency check, where the agreement between band structure and ARPES confirms the correctness of the calculations.

\section{Four-band model around the Dirac points -- group theory analysis}
\label{Sec:FourBandModel}

The 8-band model constructed above can be further simplified when only considering states close to the zero energy. These states are around the Dirac points $K=(\frac{4\pi}{\sqrt 3 a},0)$ and $K^\prime=-K$. Without loss of generality, we develop the 4-band model around K. This point is of high symmetry in the Brillouin zone, and we will fully use the lattice symmetry groups to analyze the energy band structure around it. The related group theory knowledge is reviewed in Appendix~\ref{appendix:group}. The low-energy model around the $K^\prime$ point can be directly obtained by performing the Kramers transformation.

\subsection{Gapless Dirac points in the absence of spin-orbit coupling}
Let us begin with the case in the absence of substrate and without SOC. We only need to consider the sublattice and orbital degrees of freedom of the honeycomb lattice. The rotation symmetry to maintain the K-point invariant is reduced to 3-fold. Hence, the little group symmetry for the K-point is $D_{3h}$, which possesses one 3-fold vertical rotation axis, three 2-fold horizontal rotation axes, three vertical reflection planes, and the horizontal reflection plane. For simplicity, we will use its subgroup $C_{3v}$ to explain the energy degeneracy pattern at the K-point, which is already sufficient for most discussions in this subsection.

We define the bases within the sector of the $\sigma$-orbitals $p_x$ and $p_y$ for the Bloch-wave states at K. The circularly polarized orbital states $p_\pm =\frac{1}{\sqrt 2}(p_x\pm i p_y)$ carry angular momentum, hence, they are more natural bases to manifest the point group symmetries. There are 4 states with the wavevector K denoted as $|\psi_{A+}(K)\rangle$, $|\psi_{B-}(K)\rangle$, $|\psi_{A-}(K)\rangle$, and $|\psi_{B+}(K)\rangle$ where A and B refer to two different sublattices, and $\pm$ refer to $p_\pm$-orbitals, respectively. The orbital angular momenta of these bases come from two sources. First, the $p_\pm$-orbitals carry the on-site orbital angular
momentum $L_{ob}=\pm 1$, respectively. The second contribution arises from the phase winding around each plaquette. The plane-wave phase factor on each site takes the value from $1$, $\omega=e^{i\frac{2}{3}\pi}$, and $\omega^2$. Its winding patterns around each plaquette are counter-clockwise for $|\psi_{A\pm}(K)\rangle$ and clockwise for $|\psi_{B\pm}(K)\rangle$, contributing to the angular momentum $L_{pl}=\pm 1$, respectively.

Now combine the on-site and plaquette orbital angular momenta: The total orbital angular momenta for $|\psi_{A+}\rangle$ and $|\psi_{B-}\rangle$ are $L_z=\pm 2\equiv \mp 1 (\mbox{mod}~ 3)$, respectively. Hence, they form the two-fold degenerate $E$-representation of the $C_{3v}$ group, whose energy is defined as the reference zero energy. On the other hand, $L_z=0$ for both $|\psi_{A-}(K)\rangle$ and $|\psi_{B+}(K)\rangle$. Their superpositions $\frac{1}{\sqrt 2} (|\psi_{A-}(K)\rangle \pm |\psi_{B+}(K)\rangle$ belong to the one-dimensional representations of $A_{1,2}$, which are non-degenerate. The former has a higher energy above the Dirac point, and the latter is below the Dirac point.

\subsection{The on-site spin-orbit splitting}
Now, let us consider SOC but without the effect from the substrate. We project out the states of $\frac{1}{\sqrt 2} (|\psi_{A-}(K)\rangle \pm |\psi_{B+}(K)\rangle$, since they are away from the zero energy at the order of the band width. Only the states of the $E$-representation for the orbital wave functions are kept as the low energy sector around the K-point. Taking into account the spin degeneracy, the low energy Hilbert space is 4-dimensional. It is spanned by $|\psi_{A+,\alpha}(K)\rangle=|\psi_{A+}(K)\rangle\otimes |\alpha\rangle$ and $|\psi_{B-,\alpha}(K)\rangle=|\psi_{B-}(K)\rangle\otimes |\alpha\rangle$, where $\alpha=\uparrow, \downarrow$ represents the $s_z$-eigenvalues.

In the absence of the substrate, the little group for the K-point is the $D_{3h}^D$, the double group of $D_{3h}$. The Hilbert space can be decomposed into $E_{\frac{1}{2}}$ and $E_{\frac{3}{2}}$ states, both of which are two-dimensional irreducible representations, and their angular momenta include both orbital and spin contributions. The $E_{\frac{1}{2}}$ states are spanned by
\begin{eqnarray}
E_{\frac{1}{2}}: ~|\psi_{A+,\uparrow}(K)\rangle, \ \ \, |\psi_{B-,\downarrow}(K)\rangle,
\end{eqnarray}
whose $J_z$ eigenvalues can be simply added up as $J_z=\pm\frac{5}{2}\equiv\mp\frac{1}{2} (\mbox{mod}~3)$. The $E_{\frac{3}{2}}$ states are spanned by the bases of
\begin{eqnarray}
E_{\frac{3}{2}}: ~|\psi_{A+,\downarrow}(K)\rangle, \ \ \,
|\psi_{B-,\uparrow} (K) \rangle.
\label{eq:E32}
\end{eqnarray}
These states carry different characters $\mp i$ under the horizontal reflection operation $\sigma_h$, but transform into each other under the vertical reflection operations. Although their $J_z$ eigenvalues are essentially the same as $J_z=\frac{3}{2}\equiv -\frac{3}{2}~(\mbox{mod}~3)$, they remain degenerate. In fact, it can be checked that the vertical reflections anti-commute with the horizontal, which is a special property for spinor states, which ensures their degeneracy.

In the absence of the substrate, the only spin-orbit coupling is the on-site one, as discussed before. Within the sector of the $\sigma$-orbitals, it is reduced
to $\frac{1}{2} \lambda_{\mbox{soc}}\sum_i L_z(i) \sigma_z(i)$.
Since $L_z \sigma_z$ takes value of $\pm \frac{1}{2}$ for the $E_{\frac{1}{2}}$ and $E_{\frac{3}{2}}$ sectors, respectively, the $E_\frac{1}{2}$ states are at a higher energy of $\frac{1}{2}\lambda_{\mbox{soc}}$, while the $E_{\frac{3}{2}}$ states are at a lower energy of $-\frac{1}{2}\lambda_{\mbox{soc}}$.

\subsection{The Rashba splitting due to the substrate}
Now, we consider the effect of the substrate, which breaks the horizontal reflection symmetry. The little group for the K-point is reduced to $C_{3v}^D$, the double group of $C_{3v}$. The $E_{\frac{1}{2}}$-doublet remains an irreducible representation of the $C_{3v}^D$ group, hence, their degeneracy is not affected by the substrate.

However, the $E_{\frac{3}{2}}$ sector behaves very differently. The two bases in Eq.~(\ref{eq:E32}) share the same value of $J_z$. When lacking the horizontal reflection symmetry, they are mixed and the degeneracy is lifted. According to the Rashba Hamiltonian $H_R$ constructed in Sect. \ref{sect:rashba}, under the bases of $|\psi_{A+,\downarrow}(K)\rangle$ and $|\psi_{B-,\uparrow}(K)\rangle$, it is expressed by a $2\times 2$ matrix as
\begin{eqnarray}
H_R(K)=\frac{3}{2}\lambda_R \left(
\begin{array}{cc}
 0& -i\\
 i& 0
\end{array}
\right).
\end{eqnarray}
The eigenstates are reorganized as
\begin{eqnarray}
|\psi_{\frac{3}{2},\pm i}(K)\rangle=
\frac{1}{\sqrt 2} \Big\{ |\psi_{A+,\downarrow}(K)\rangle \mp i
|\psi_{B-,\uparrow} (K) \rangle \Big\},
\end{eqnarray}
with the energy splitting $\Delta E=3 \lambda_R$. In other words, the $E_{\frac{3}{2}}$ sector splits into two non-equivalent one-dimensional representations. They are eigenstates for the vertical reflection operations, say, $\sigma_{xz}$ with respect to the $xz$-plane with the eigenvalues of $\pm i$, respectively.

\subsection{The 4-band Hamiltonian around the K-point}
\label{sect:rashba}
Having explained the degeneracy pattern at the K-point, we are ready to present the low energy Hamiltonian around the K-point. We will project the Hamiltonian $H_0+H_{\mbox{soc}}+H_R$ in Eq. (1), into the low energy bases of $|\psi_{A+,\uparrow}\rangle, |\psi_{B-,\downarrow}\rangle, |\psi_{A+,\downarrow}\rangle$, and $|\psi_{B-,\uparrow}\rangle$ .

The Rashba spin-orbit coupling $H_R$ is due to the breaking of the horizontal reflection symmetry. Based on symmetry analysis, $H_R$ is constructed as
\begin{eqnarray}
 H_R=i\lambda_R \sum_{i\in A, j}
 \big\{ c^\dagger_{i+\hat a_j, \vec p \cdot \hat a_j}
 (\hat d_j \cdot \vec \sigma) c_{i,\vec p\cdot \hat a_j} +h.c.\big\},
\end{eqnarray}
where $i$ is the site index of the $A$-sublattice, $\hat a_j$'s with $j=1\sim 3$ represents the unit vectors along the nearest neighboring bonds, $\vec p \cdot \hat a_j$ is the $p$-orbital along the $\hat a_j$-bond direction, and $\hat d_j=\hat z \times \hat a_j$ is the Dzyaloshinskii-Moriya vector along the $\hat a_j$-bond.

We define the 4-component spinor around the K-point,
\begin{eqnarray}
\Psi(\vec K+\vec q)&=& \Big (|\psi_{A+,\uparrow}
(\vec K+\vec q)\rangle,
|\psi_{B-,\downarrow} (\vec K+\vec q)\rangle, \nonumber \\
&&|\psi_{A+,\downarrow} (\vec K+\vec q)\rangle,
|\psi_{B-,\uparrow} (\vec K+\vec q)\rangle
\Big )^T. ~~
\end{eqnarray}
The 4-band Hamiltonian is expressed as $H=\sum_q \Psi^\dagger (\vec K+ \vec q) H(\vec q) \Psi(\vec K+ \vec q)$, where $\vec q$ is the small deviation from the K-point. The matrix kernel $H(\vec q)$ is a $4\times 4$ matrix expressed as
\begin{eqnarray}
H(\vec q)=\left(
\begin{array}{cc}
A& B\\
B^\dagger &C
\end{array}
\right).
\label{eq:4band}
\end{eqnarray}
$A$, $B$ and $C$ are the $2\times 2$ block matrices defined as
\begin{eqnarray}
A&=&\left(
\begin{array}{cc}
\lambda_{soc} & -i \lambda_R f_2(\vec q)\\
i \lambda_R f_2^*(\vec q)& \lambda_{soc}
\end{array}
\right), \nonumber \\
B&=&t_\parallel \left(
\begin{array}{cc}
 0& f_0(\vec q) \\
 f_0^*(\vec q) &0\\
\end{array}
\right), \\
C&=&\left(
\begin{array}{cc}
-\lambda_{soc} & -i \lambda_R f_1(\vec q)\\
i \lambda_R f_1^*(\vec q)& -\lambda_{soc}
\end{array}
\right),
\end{eqnarray}
where $f_k(\vec q)$ with $k=0,1,2$ are defined as
\begin{eqnarray}
f_k(\vec q)&=&\sum_{j=1}^3 \omega^k e^{i(\vec q \cdot \hat a_j-2 \theta_j)}, \end{eqnarray}
with $\theta_j=\frac{2\pi}{3}j-\frac{\pi}{2}$ the azimuthal angle for the bond orientation of $\hat a_j$. Expand $H(\vec q)$ at small values of $q$, we arrive at
\begin{eqnarray}
 H(\vec q)=\left(
\begin{array}{cccc}
\lambda_{soc}& -\frac{3i}{4}\lambda_R q_+ & 0 & -\frac{3}{4} t_\parallel q_-\\
\frac{3i}{4}\lambda_R q_- & \lambda_{soc}& -\frac{3}{4} t_\parallel q_+& 0 \\
0& -\frac{3}{4}t_\parallel q_- & -\lambda_{soc} & -\frac{3}{2}i \lambda_R \\
-\frac{3}{4}t_\parallel q_+& 0 & \frac{3}{2} i \lambda_R &-\lambda_{soc}
\end{array}
\right),
\end{eqnarray}
where $q_\pm =q_x\pm i q_y$.

Similarly, at $K^\prime=-K$, the $E_{\frac{1}{2}}$-doublet becomes $|\psi_{B+,\uparrow}(K^\prime)\rangle$ and $|\psi_{A-,\downarrow}(K^\prime)\rangle$, and the $E_{\frac{3}{2}}$ doublet becomes $|\psi_{B+,\downarrow} (\vec K^\prime)\rangle$, and $|\psi_{A-,\uparrow} (\vec K^\prime)\rangle$. We define the 4-component spinor around $K^\prime$,
\begin{eqnarray}
\Psi(\vec K^\prime+\vec q)&=& \Big (|\psi_{B+,\uparrow}
(\vec K^\prime+\vec q)\rangle,
|\psi_{A-,\downarrow} (\vec K^\prime+\vec q)\rangle, \nonumber \\
&&|\psi_{B+,\downarrow} (\vec K^\prime+\vec q)\rangle,
|\psi_{A-,\uparrow} (\vec K^\prime+\vec q)\rangle
\Big )^T, ~~~~
\end{eqnarray}
where $\vec q$ is the deviation from the $K^\prime$-point. By performing the Kramers' transformation, the 4-band Hamiltonian around $K^\prime$ is expressed as
$H^\prime=\sum_q \Psi^\dagger (\vec K^\prime+ \vec q) H^\prime(\vec q) \Psi(\vec K^\prime+ \vec q)$ with the matrix kernel $H^\prime(\vec q)$ expanded at small values of $q$ as
\begin{eqnarray}
 H^\prime(\vec q)=\left(
\begin{array}{cccc}
 \lambda_{soc}& -\frac{3i}{4}\lambda_R q_+ & 0 & \frac{3}{4} t_\parallel q_-\\
 \frac{3i}{4}\lambda_R q_- & \lambda_{soc}& \frac{3}{4} t_\parallel q_+& 0 \\
 0& \frac{3}{4}t_\parallel q_- & -\lambda_{soc} & \frac{3}{2}i \lambda_R \\
 \frac{3}{4}t_\parallel q_+& 0 & -\frac{3}{2} i \lambda_R &-\lambda_{soc}
\end{array}
\right).
\end{eqnarray}

\section{Summary and Outlook}
\label{Sec:Summary}
2D TIs such as QSH insulators have a natural advantage over their 3D cousins, in that the edge states of a QSH insulators are more robust against non-magnetic scattering, because the only possible backscattering channel is forbidden. So far, most theoretical studies rely on free-standing films, the chemical stability of which is usually very poor. Thus, it is natural to place the film on a substrate, but clearly the electronic and, in particular, topological properties of a free-standing layer will most likely be affected by the substrate. Thus, a primary aim is to search for large-gap QSH states existing on a monolayer plus substrate system.
The concrete downfolding of our (Bi,As,Sb)/SiC systems on an effective low-energy Hamiltonian description reveals a cornerstone for the paradigm, where the substrate stabilizes the monolayer on the one hand (pushes p$_x$ and p$_y$ orbitals to the Fermi level)  but on the other hand allows for the on-site SOC, creating a large topological band gap in our Bi/SiC-system of order $\sim0.7$ eV.

We emphasize, that in our experience it is the interplay of theory and experiment as it has crystallized in this work, which provides a kind of "smoking gun" logical argument for the paradigm:
(i) Firstly, as shown in Fig.~3, the theoretical band structure, based on an a-priori DFT calculation, including SOC and the ARPES data (here for Bi/SiC) display a close overall agreement, and a particularly good match around the topologically most relevant K-points in the BZ.
(ii) It is this band structure which, in a step-by-step comparison with other key data from experiment (such as the STM-derived large band gap, etc. ), then confirms the topological electronic structure via an explicit evaluation of the topological constant $Z_2$ (Appendix~\ref{appendix:TopInv}). Of course, imminent transport experiments proving the edge-current quantization are of utmost importance.\cite{Dominguez2018} 
(iii) As shown in detail in the present work, when "downfolded" to the effective low-energy Hamiltonian in Eq.~(1), the internal consistency of our arguments can be further illustrated and checked. The Rashba term $\sim\lambda_\text{R}$ can directly be seen in ARPES data as a valence-band splitting and the local (on-site) SOC term $~\lambda_{SOC}$ is responsible for the large bulk gap and is seen in STM data\cite{Reis2017:S} and, of course, should be detectable also in optical absorption data.

A possible extension and application of our low-energy Hamiltonian in Eq. (\ref{EffLowEnergyHamiltonianGeneral}) concerns the quantum anomalous Hall (QAH) insulators with large gaps. The integer quantum Hall effect\cite{klitzing-80prl494} was the first experimentally realized topological state of matter in 2D, which arises in quasi-2D electron gases, in magnetic fields with integer fillings of Landau levels. In the quantum Hall effect, the quantization of the Hall conductance is protected by the nontrivial band structure topology characterized by the Chern number.\cite{thouless82-prl405} In order to achieve a non-zero Chern number pattern, time--reversal symmetry needs to be broken, but Landau levels are not necessary. Haldane,\cite{haldane88prl2015} early on, has constructed a model for QAH states, {\it i.e.} a tight-binding model in the honeycomb lattice with Bloch wave structures, and showed that it carries quantum Hall states with $\nu=\pm1$. This effect is termed QAH effect, because the net magnetic flux is zero in each unit cell, and there are no Landau levels.
In our Sec.~\ref{Sec:FourBandModel} above, we constructed a minimal (4-band, $p_x$- and $p_y$-orbitals only) model in the honeycomb lattice. We studied the conditions for achieving the QAH insulator, further simplified within a model, keeping the inversion symmetry ($\perp z$).\cite{ZhangGF2014,Zhang2011:PRA} The first idea is again, that the multi-orbital ($p_x$,$p_y$) structure allows for the atomic SOC (Sec.~\ref{Sec:FourBandModel}). This, as a consequence, lifts the degeneracy between two sets of on-site Kramers doublets, {\it i.e.} $j_z=\pm3/2$ and $j_z=\pm1/2$. Alternatively to our derivation in Sec.~\ref{Sec:ElectronicStructure}, one can already in the very first, {\it i.e.} on-site step, involve the atomic SOC coupling $\bm{L}\cdot\bm{S}$ on each site. This amounts to work with the eigenstates $p_{\pm,s=\uparrow,\downarrow}^{\dagger}=(p_{x,s}^{\dagger}\pm\i p_{y,s}^{\dagger})/\sqrt{2}$, which are the orbital angular momentum $L_z$ eigenstates (and $j_z$ is the $z$ component of the total angular momentum; for details see Ref.~\onlinecite{ZhangGF2014}). Clearly, the $p_{\pm,s}^{\dagger}$-basis has already encoded the topological properties of the left- (spin up) and right-moving (spin down) edge currents in the TRI situation of the QSH systems.
In the present work, our minimal model (Sec.~\ref{Sec:FourBandModel}) is extended to a  more general "down-folded" Hamiltonian for the (Bi,As,Sb)/SiC systems. A new objective (by including a TRI symmetry-breaking Neel exchange term) in future work will be to make predictions for QAH insulators with large gaps in concrete realizations of monolayer-substrate systems.

\acknowledgements
We acknowledge financial support from the DFG via SFB 1170 "ToCoTronics", DFG-SPP 1666, ERC-StG-TOPOLECTRICS, and the ENB Graduate School on Topological Insulators.
G. L. acknowledges the starting grant of ShanghaiTech University and Program for Professor of Special Appointment (Shanghai Eastern Scholar) while C. W. is supported by AFOSR FA9550-14-1-0168. Calculations were carried out at the Leibniz Supercomputing Centre (LRZ) in Munich, and the HPC Platform of Shanghaitech University Library and Information Services. 

\appendix

\section{DFT Details}
\label{appendix:DFTdetails}

Throughout our calculations, the projector augmented wave pseudopotential~\cite{PhysRevB.50.17953} was employed, and the exchange-correlation energy was treated with the generalized gradient approximation (GGA) of Perdew, Burke and Ernzerhof (PBE)~\cite{perdew1996}. The spin-orbit coupling of electrons was considered self-consistently in our calculations. The cutoff energy of the plane-wave basis functions was set to be 500 eV. For the reciprocal-space integration we used the Monkhorst-Pack special k-point method~\cite{PhysRevB.13.5188} with 9x9x1 grid. The energy convergence criteria was set to be 1meV/atom. The atomic positions were fully relaxed using the conjugate gradient algorithm until all interatomic forces were small than 0.01 eV/A. To calculate the ${\cal Z}_{2}$ topological invariant, we constructed the tight-binding (TB) model Hamiltonian according to the results of DFT band structure from the maximally localized Wannier functions (MLWFs)~\cite{Marzari1997} by using the VASP2WANNIER90 interface~\cite{Mostofi2008}. 

\subsection{Geometry of freestanding monolayer Sb/As}

\begin{figure}[ht]
\centering
\includegraphics[width=\linewidth]{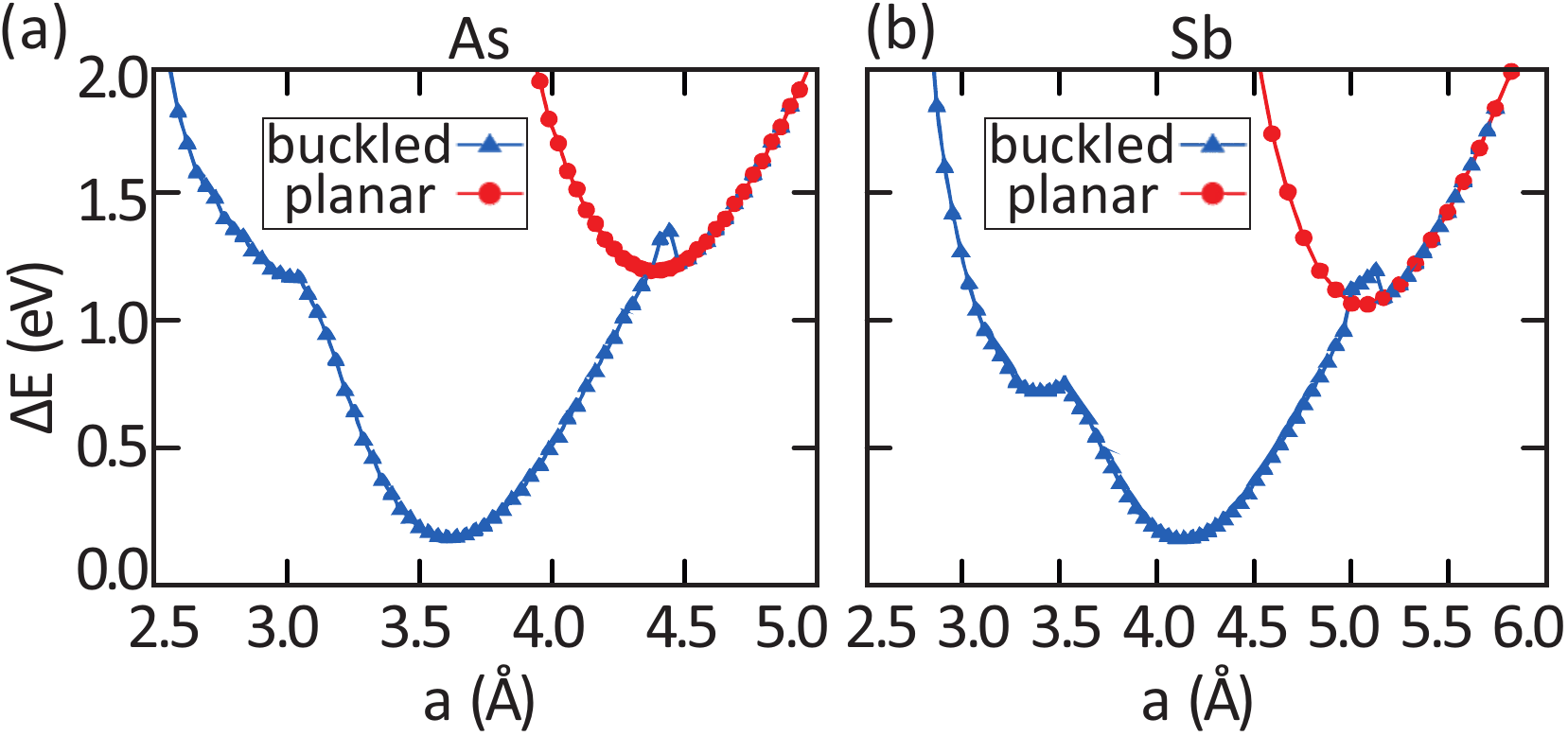}
\caption{Total energy of the buckled (in blue) and planar (in red) honeycomb lattice of (a) arsenic and (b) antimony systems. In both plots the energy zero is set to the energy of the optimal structure.}
\label{optimization}
\end{figure}

Honeycomb layers formed by group-V elements can be in either buckled or in planar forms, with both having threefold rotation and inversion symmetry, similar to graphene. 
Structure optimizations (DFT) indicate that both arsenic and antimony prefer a buckled honeycomb lattice, as shown in Fig.~\ref{optimization}, where
the total energy relative to that of the optimized structure is displayed for both systems. Here, the blue and red lines correspond to the buckled and planar configurations, respectively. 
Overall, the buckled structure is more stable than the planar configuration in both systems, with the optimal lattice constants being $a=3.605$ \AA ~for arsenic and $a=4.121$ \AA ~for antimony, respectively.
Though the energy barrier of the planar configuration is of the order of 1 eV in both systems, they can be locally stabilized. To achieve this, the lattice constants need to significantly be enlarged to $a=4.375$ \AA ~and $a=5.084$ \AA ~in the two systems. 
For the antimony planar honeycomb, this means a tensile strain of $4.972\%$ will be imposed if SiC(0001) is chosen as a substrate, which is still of experimental operating feasibility.    
As in the case of bismuthene, SiC stabilizes the planar configuration of the overlaid antimony layer by bonding with it yet keeping its $sp^2$ in plane configuration.

Due to the reduced lattice size compared to bismuthene, it is of experimental interest to search for other insulating substrates more compatible to the optimal lattice constants of arsenic and antimony, which is essential for achieving the largest topological gap allowed in each system. The constraint on such substrate materials is the same as that for bismuthene, {\it {\it i.e.}} they shall keep the planar form of arsenic and antimony lattices, yet not altering their low-energy electronic structures. The aim of this paper is to demonstrate the general principle for achieving the large topological gap and illustrate its universality in heavy-atom honeycomb/substrate combinations. Thus, we will not discuss other substrates but keep the following discussion on As/SiC and Sb/SiC, assuming in both cases a planar layer. 

\subsection{\texorpdfstring{Electronic structure and ${Z}_{2}$ invariant}{Electronic structure and \textit{Z}2 invariant}}

As a result of the lattice planar configuration, the electronic structures of the group-V arsenic and antimony layers on SiC(0001) (see Fig.~\ref{SbAs_band}) highly resemble that of bismuthene. 
The characteristic features of these band structures appearing at the K point of the Brillouin zone include: (1) a linear band crossing in the absence of SOC, (2) the opening of topological gaps for finite SOC, and (3) the Rashba splitting of the top valence band after including the SOC.

Point (1) appears as a result of the honeycomb lattice structure, similar to graphene. Points (2) and (3) result from the combined effect of SOC and the inversion symmetry breaking, as thoroughly explained in bismuthene~\cite{Reis2017:S} and in the Introduction of this paper.

{As in Bi/SiC, the low-energy effective model of As/SiC and Sb/SiC at the K point is governed by the $p_{x/y}$ orbitals. As discussed before, the $p_{z}$-orbital (As,Sb) component in the combined layer/substrate systems is shifted to higher energies (and is particularly strong around the $\Gamma$-point). It creates the Rashba splitting of the top valence band by coupling by coupling the potential gradient, created by the substrate, to the $p_{x/y}$ orbitals (see Sec.~\ref{Sec:LowEnergyModel}). The only difference to bismuthene lies in the reduced atomic SOC strength of arsenic and antimony. Thus, the size of the topological gap and the valence-band Rashba splitting shrinks correspondingly.}

\section{\texorpdfstring{The $C_{ 3V}$ group and its double group $C_{3V}^D$}{The C3V group and its double group C3VD}}
\label{appendix:group}
The $C_{3v}$ group includes 6 operations in 3 conjugacy classes:
the identity I, the 3-fold rotations
$\{C_3^1, C_3^2\}$ around the vertical axis, and the
reflection operations with respect to three vertical planes
$\{\sigma_{v_i}\}$ with $i=1\sim 3$.
It possesses two one-dimensional representations $A_1$ and $A_2$,
and one two-dimensional representation $E$.
Their character table is presented in Tab \ref{table_c3v}.
The bases of the $A_{1,2}$ representations carry angular momentum
quantum number $L_z=0$, and those of the $E$ representation
can be chosen with $L_z=\pm 1$.

\begin{table}[ht]
\begin{center}
\begin{tabular}{|c|l|c|c|c|} \hline
 & I & 2$C_3$ & 3$\sigma_v$ \\ \hline
$A_1$& 1& 1& 1 \\ \hline
$A_2$& 1& 1& -1 \\ \hline
$E$ & 2& -1& 0 \\ \hline
\end{tabular}
\caption{The character table of the $C_{3v}$ group, which has
two one dimensional representations $A_{1,2}$ and one two-dimensional
representation $E$.
$A_{1,2}$ carry orbital angular momentum $L_z=0$, and
$E$ carries $L_z=\pm 1$.
}
\end{center}
\label{table_c3v}
\end{table}

In the presence of spin-orbit coupling, $C_{3v}$ is augmented to its
double group $C_{3v}^D=C_{3v} + \bar C_{3v}$.
$\bar C_{3v}=\bar I C_{3v}$ is the coset by multiplying $\bar I$
to $C_{3v}$, where $\bar I$ is the rotation of $2\pi$.
The $C^D_{3v}$ group has 6 conjugacy classes, and hence 6 non-equivalent
irreducible representations whose characteristic table is
presented in Tab. \ref{table_c3vd}.
$A_{1,2}$ and $E$ remain the representations of $C_{3v}^D$ of integer
angular momentum, for which $\bar I$ is the same as the identity
operation.
In addition, $C^D_{3V}$ also possesses half-integer angular momentum
representations, for which $\bar I$ is represented as the negative
of identity matrix.
For example, a new two-dimensional representation $E_\frac{1}{2}$ appears
corresponding to the angular momentum $J_z=\pm\frac{1}{2}$.
The cases of $J_z=\pm\frac{3}{2}$ are often denoted as the
$E_{\frac{3}{2}}$ representation.
Actually they are not an irreducible two-dimensional representation, but
two non-equivalent one-dimensional representations.
The two bases of $\psi_{J_z=\pm\frac{3}{2}}$ are equivalent under the 3-fold
rotations since $\frac{3}{2}\equiv -\frac{3}{2} (\mbox{mod}~ 3)$, and neither
of them are eigenstates of the reflections $\sigma_v$ and
$\bar\sigma_v=\bar I \sigma_v$.
Instead, their superpositions $\frac{1}{\sqrt 2} (\psi_\frac{3}{2}\pm
i\psi_{-\frac{3}{2}})$ carry the characters of $\pm i$ for $\sigma_v$
and $\mp i$ for $\bar \sigma_v$, respectively.

\begin{table}[htbp]
\begin{center}
\begin{tabular}{|c|c|c|c|c|c|c|} \hline
& I & $\bar I$ & $\{C^1_3, \bar C^2_3$ \} & $\{C^2_3, \bar C^1_3\}$ &
 3$\sigma_v$ & 3$\bar \sigma_v$ \\ \hline
$E_{\frac{1}{2}}$ & 2 &-2& 1 & -1 & 0 &0 \\ \hline
$E_{\frac{3}{2}}$ &1 &-1& -1 & 1 & $i$ & $-i$ \\
 &1 &-1& -1 & 1 &$-i$ & $i$ \\ \hline
\end{tabular}
\caption{Spinor representations for the $C_{3v}^D$ group:
The two-dimensional representation $E_{\frac{1}{2}}$ is of $J_z=\pm \frac{1}{2}$.
$E_{\frac{3}{2}}$ splits into two non-equivalent one-dimensional
representations with different characters under vertical
reflections.
}
\end{center}
\label{table_c3vd}
\end{table}

\section{Topological Invariant}
\label{appendix:TopInv}

Following the same recipe as for bismuthene, the characterization of the QSH phase can be equally done for arsenic and antimony layers (arsenene and antimonene, respectively), by constructing a slab of their honeycomb lattices with either zigzag or armchair edges. Inside the topological gap, there appear states connecting the bulk valence and conduction bands, spatially residing at the edges. Alternative to this, as described here, we follow the bulk-boundary correspondence to calculate the $Z_{2}$ invariant from their bulk electronic structures. 
The topology of the tight-binding model is inherently determined by the Berry curvature of the occupied bands, which can be extracted from two different strategies.

\begin{figure}[htbp]
\centering
\includegraphics[width=\linewidth]{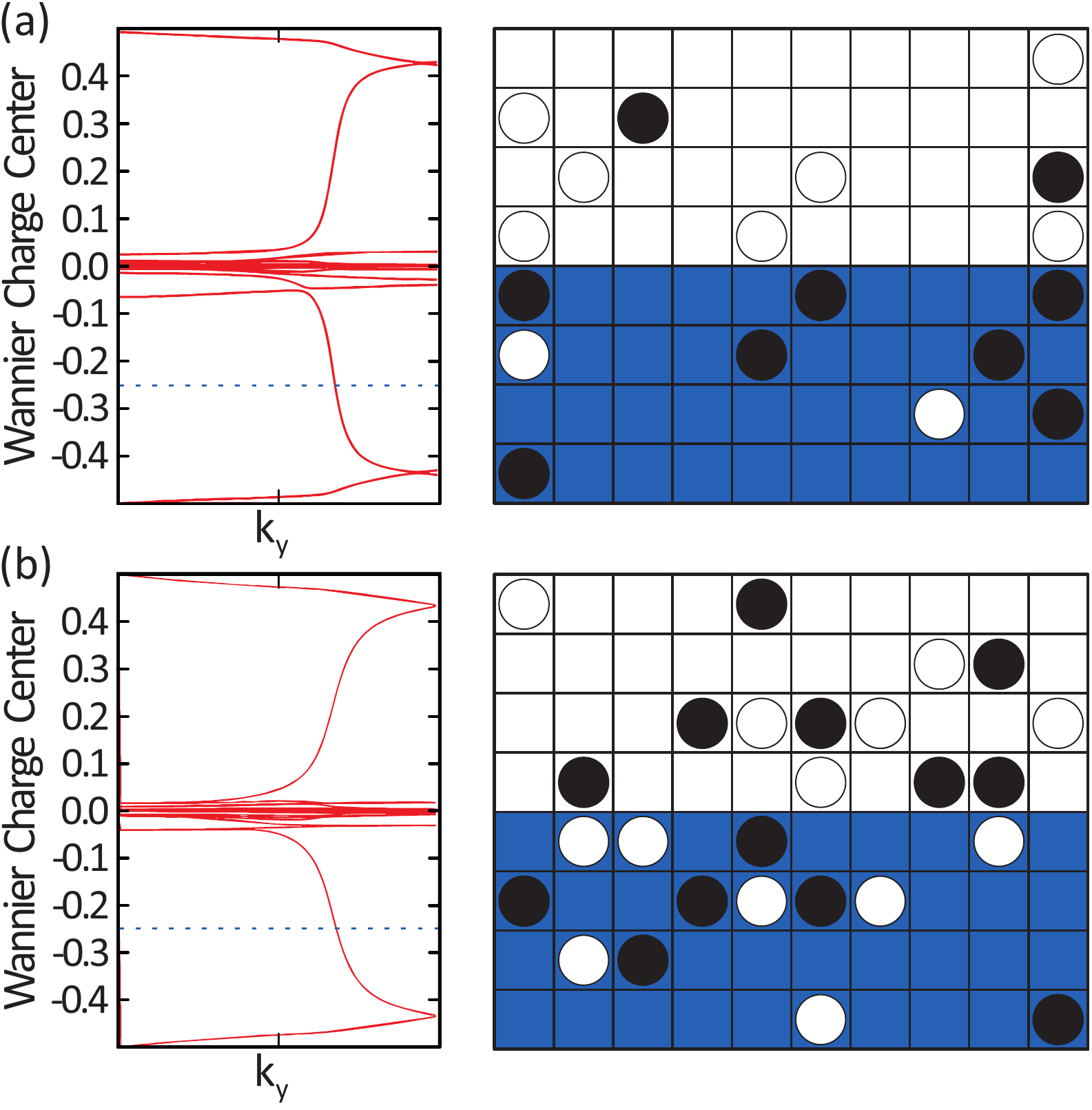}
\caption{Topological characterization from Wilson loop (left column) and topological obstruction (right column) for (a) arsenene and (b) antimonene, respectively.}
\label{Z2}
\end{figure}

In Fig.~\ref{Z2}, the corresponding Wilson loop and the topological obstruction plots are shown \cite{Vanderbilt2011}. The Wilson loop traces the change of the Wannier charge center along a closed path in parameter space (here it is the momentum). Ordinary insulators have logarithmically localized orbitals with their Wannier charge center (shown as red solid lines in Fig.~\ref{Z2}) nearly constant in momentum space. In contrast, the QSH states have nontrivial Berry curvature structure in the entire half BZ. As a result, the Wannier charge center switches from one to another following a closed path. 
In our calculations this is reflected from the odd number of crosses of a straight line (blue dashed line in Fig.~\ref{Z2}) with the Wannier charge center, see Fig.~\ref{Z2} for the case of arsenene and antimonene.

Topological obstruction is another bulk topological characterization. 
It is based on the fact that, for a topological nontrivial system, it is not possible to smoothly define a gauge for the Bloch wavefunctions along a closed path in half of the BZ. The change of Berry phase must be an odd integer times $2\pi$. 
Numerically, we discretized half of the BZ into small parquets, along the boundary of which the Berry connections were calculated. The sum of them over all parquets modulo $2\pi$ gives rise to the $Z_{2}$ number. In each parquet this number can either $0$ or $\pm 1$, denoted by empty, empty circles and solid circles in the chess plot of Fig.~\ref{Z2}. One can immediately see that both methods demonstrate that these two systems, {\it {\it i.e.}}  arsenene and antimonene, are indeed topologically nontrivial. 
We note that the two methods used in the current work are equivalent to the characterization via the edge states that we employed for bismuthene~\cite{Reis2017:S}.

\section{Detailed derivation of the effective low-energy model}
\label{appendix:LowEnergyModel}
The direct and reciprocal lattice vectors of the honeycomb layer are given as:
\begin{eqnarray}
\mathbf{a_{1}} &=& \hat{x}\;, \hspace{2cm} \mathbf{a_{2}} = -\frac{1}{2}\hat{x} + \frac{\sqrt{3}}{2}\hat{y}\;,\cr
\mathbf{b_{1}} &=& (2\pi, \frac{2\sqrt{3}}{3}\pi)\;, \hspace{0.3cm} \mathbf{b_{2}} = (0, \frac{4\sqrt{3}}{3}\pi)\;.
\end{eqnarray}
The high-symmetry points are $\Gamma=(0, 0)$, $M=(\pi, \frac{\sqrt{3}}{3}\pi)$, $K=(\frac{2\pi}{3}, \frac{2\sqrt{3}}{3}\pi)$, and $K^{\prime}=(\frac{4\pi}{3}, 0)$.

The basis functions of the $\sigma$-bands model are taken as 
\begin{equation}
|p_{x\uparrow}^{A}\rangle, |p_{y\uparrow}^{A}\rangle, |p_{x\uparrow}^{B}\rangle, |p_{y\uparrow}^{B}\rangle; \hspace{0.5cm}
|p_{x\downarrow}^{A}\rangle, |p_{y\downarrow}^{A}\rangle, |p_{x\downarrow}^{B}\rangle, |p_{y\downarrow}^{B}\rangle\;.
\end{equation}
A and B indicate the two inequivalent sites in one unit-cell of the honeycomb lattice. 
The Hamiltonian matrix elements in two spin sectors are equivalent. It is straight forward to get them from Slater-Koster integrals,
\begin{equation}\label{H1}
H^{\sigma\sigma}_{\uparrow\uparrow}=H^{\sigma\sigma}_{\downarrow\downarrow} =
\begin{pmatrix}
h_{xx}^{AA} & 0 & h_{xx}^{AB} & h_{xy}^{AB} \cr
0 & h_{yy}^{AA} & h_{yx}^{AB} & h_{yy}^{AB} \cr
\dagger  & \dagger & h_{xx}^{BB} & 0 \cr
\dagger &  \dagger & 0 & h_{yy}^{BB}
\end{pmatrix}\;,
\end{equation}
with 
\begin{subequations}\label{SK_integral}
\begin{align}
& h_{xx}^{AA} = h_{xx}^{BB} = V_{pp\sigma}^{0}\;, \\
& h_{yy}^{AA}=h_{yy}^{BB}=V_{pp\pi}^{0}\;, \\
& h_{xx}^{AB}=V_{pp\pi}^{1}+\frac{1}{2}(3V_{pp\sigma}^{1}+V_{pp\pi}^{1})e^{i\frac{\sqrt{3}}{2}k_{y}}\cos\frac{k_{x}}{2}\;, \\
& h_{xy}^{AB}=h_{yx}^{AB}=i\frac{\sqrt{3}}{2}(V_{pp\sigma}^{1}-V_{pp\pi}^{1})e^{i\frac{\sqrt{3}}{2}k_{y}}\sin\frac{k_{x}}{2}\;, \\
& h_{yy}^{AB}=V_{pp\sigma}^{1}+\frac{1}{2}(V_{pp\sigma}^{1}+3V_{pp\pi}^{1})e^{i\frac{\sqrt{3}}{2}k_{y}}\cos\frac{k_{x}}{2}\;.
\end{align}
\end{subequations}
Diagonalizing the Hamiltonian at K-point yields the following four eigenenergies
\begin{subequations}
\begin{align}
E_{1/2}&=&\frac{1}{4}\bigg[2(V_{pp\sigma}^{0}+V_{pp\pi}^{0}) +3(V_{pp\pi}^{1}-V_{pp\sigma}^{1})\nonumber\\
&&\pm \sqrt{4(V_{pp\sigma}^{0}-V_{pp\pi}^{0})^{2}+9(V_{pp\pi}^{1}-V_{pp\sigma}^{1})^{2}}\bigg]\;,\\
E_{3/4}&=&\frac{1}{4}\bigg[2(V_{pp\sigma}^{0}+V_{pp\pi}^{0})-3(V_{pp\pi}^{1}-V_{pp\sigma}^{1})\nonumber\\
&&\pm \sqrt{4(V_{pp\sigma}^{0}-V_{pp\pi}^{0})^{2}+9(V_{pp\pi}^{1}-V_{pp\sigma}^{1})^{2}}\bigg]\;.
\end{align}
\end{subequations}
Two of the four states are degenerate at the K-point and stay right at the Fermi level, which constrains parameters $V_{pp\sigma}^{0}, V_{pp\pi}^{0}$ to be 0. Hamiltonian (\ref{H1}) now becomes
\begin{equation}\label{H_ss}
H_{\uparrow\uparrow}=H_{\downarrow\downarrow} = 
\begin{pmatrix}
0 & 0 & h_{xx}^{AB} & h_{xy}^{AB} \cr
0 & 0 & h_{yx}^{AB} & h_{yy}^{AB} \cr
\dagger & \dagger & 0 & 0 \cr
 \dagger & \dagger & 0 & 0
\end{pmatrix}\;,
\end{equation} 
with eigen-energies $E = 0, 0, \pm\frac{3}{2}(V_{pp\pi}^{1}-V_{pp\sigma}^{1})$ at the K-point. 

Now we proceed to include SOC. 
There are two different types of SOC in bismuthene/SiC. One is the intrinsic SOC from the induced magnetic filed due to the electron motion in the atomic electric field. 
The other one is the Rashba type SOC from the inversion symmetry breaking, due to the presence of a substrate. 
It is well known that, for $\pi$-bands, these two SOC take the following form:
\begin{equation}
H_{I} = i\lambda_{I}\sum_{\langle\langle i,j\rangle\rangle, \alpha\beta} c_{i\alpha}^{\dagger}\nu_{ij}\sigma_{\alpha\beta}^{z}c_{j\beta}
\end{equation} 
for the intrinsic SOC, and
\begin{equation}
H_{R} = i\lambda_{R}\sum_{\langle i,j\rangle, \alpha\beta}c_{i\alpha}^{\dagger}({\bm\sigma}\times \hat{d}_{i,j})\cdot\hat{z}c_{j\beta}
\end{equation}
for the Rashba SOC. 
$\nu_{ij} = +(-)$ if the hopping is (anti) clock-wise inside the haxagon.
A clear difference between the two types of SOC is that the Rashba SOC couples the different spin components whereas the intrinsic SOC does not.

However, in (Bi, Sb, As)/SiC systems the active orbitals are $p_{x}$ and $p_{y}$, and the above SOC term for $\pi$-bands does not apply. We, thus, need to get the corresponding SOC for the $\sigma$-bands. To do so, we first expand the basis function to include $s$ and $p_{z}$ orbitals as well, {\it i.e.} 
\begin{eqnarray}
&&|p_{x\uparrow}^{A}\rangle, |p_{y\uparrow}^{A}\rangle, |p_{x\uparrow}^{B}\rangle, |p_{y\uparrow}^{B},
|p_{x\downarrow}^{A}\rangle, |p_{y\downarrow}^{A}\rangle, |p_{x\downarrow}^{B}\rangle, |p_{y\downarrow}^{B}\rangle; \nonumber\\
&&|p_{z\uparrow}^{A}\rangle, |p_{z\uparrow}^{B}\rangle, |s_{\uparrow}^{A}\rangle, |s_{\uparrow}^{B}\rangle, 
|p_{z\downarrow}^{A}\rangle, |p_{z\downarrow}^{B}\rangle, |s_{\downarrow}^{A}\rangle, |s_{\downarrow}^{B}\rangle\;.
\end{eqnarray}
The Hamiltonian is a $16\times16$ matrix, and can be cast into the following form
\begin{equation}\label{Eq:Hss_pp}
H = 
\begin{pmatrix}
H^{\sigma\sigma} & H^{\sigma\pi} \cr
H^{\pi\sigma} & H^{\pi\pi} 
\end{pmatrix}\;.
\end{equation}
Without SOC, $H^{\sigma\sigma}_{\uparrow\uparrow}=H^{\sigma\sigma}_{\downarrow\downarrow}$ are given in Eq.~(\ref{H_ss}).  $H_{\pi\pi}$ is the corresponding Hamiltonian matrix spanning on the basis functions of $p_{z}$ and $s$:
\begin{equation}
H^{\pi\pi}_{\uparrow\uparrow} = H^{\pi\pi}_{\downarrow\downarrow} = 
\begin{pmatrix}
h_{zz}^{AA} & h_{zz}^{AB} & 0 & 0 \cr
h_{zz}^{BA} & h_{zz}^{BB} & 0 & 0 \cr
0 & 0 & 0 & h_{ss}^{AB} \cr
0 & 0 & h_{ss}^{AB} & 0 
\end{pmatrix}\;,
\end{equation}
where $h_{zz}^{AA}=h_{zz}^{BB}=V_{pp\pi}^{\prime}$, $h_{zz}^{AB}=V_{pp\pi}^{\prime}[1+2\cos\frac{k_{x}}{2}e^{i\frac{\sqrt{3}}{2}k_{y}}]$ and $h_{ss}^{AB}=V_{ss\sigma}[1+2\cos\frac{k_{x}}{2}e^{i\frac{\sqrt{3}}{2}k_{y}}]$ and $h_{zz}^{BA}=(h_{zz}^{AB})^{\dagger}$, 
$h_{ss}^{BA}=(h_{ss}^{AB})^{\dagger}$. 

$H^{\sigma\pi}=(H^{\pi\sigma})^{\dagger}$ in Eq.~(\ref{Eq:Hss_pp}) are the crossing terms that couple the low-energy $\sigma$-bands and the high-energy $\pi$-bands: 
\begin{equation}\label{H_sp}
H^{\sigma\pi}_{\uparrow\uparrow}=(H^{\pi\sigma}_{\uparrow\uparrow})^{\dagger}=
\begin{blockarray}{ccccc}
& \matindex{|p_{z}^{A}\rangle} & \matindex{|p_{z}^{B}\rangle} & \matindex{|s^{A}\rangle} & \matindex{|s^{B}\rangle}\\
    \begin{block}{c(cccc)}
    \matindex{|p_{x}^{A}\rangle}~ &  0 & 0 &  0 & h_{xs}^{AB}\\
    \matindex{|p_{y}^{A}\rangle}~ &  0 & 0 &  0 & h_{ys}^{AB} \\
    \matindex{|p_{x}^{B}\rangle}~ & 0 & 0 & h_{xs}^{BA} &  0 \\
    \matindex{|p_{y}^{B}\rangle}~ & 0 & 0 & h_{ys}^{BA} & 0 \\
    \end{block}
  \end{blockarray}\;,
\end{equation} 
with $h_{xs}^{AB}=\sqrt{3}iV_{sp\sigma}^{1}e^{i\frac{\sqrt{3}}{2}k_{y}}\sin\frac{k_{x}}{2}$, $h_{ys}^{AB}=-V_{sp\sigma}^{1}[1-e^{i\frac{\sqrt{3}}{2}k_{y}}\cos\frac{k_{x}}{2}]$, $h_{xs}^{BA}=-(h_{xs}^{AB})^{\dagger}=\sqrt{3}iV_{sp\sigma}^{1}e^{-i\frac{\sqrt{3}}{2}k_{y}}\sin\frac{k_{x}}{2}$, $h_{ys}^{BA}=-(h_{ys}^{AB})^{\dagger}=V_{sp\sigma}^{1}[1-e^{-i\frac{\sqrt{3}}{2}k_{y}}\cos\frac{k_{x}}{2}]$. $H_{\downarrow\downarrow}^{\sigma\pi}=(H_{\downarrow\downarrow}^{\pi\sigma})^{\dagger}$ takes the same form.

To obtain the additional matrix elements arising from $\lambda_{so}\vec{L}\cdot\vec{S}$, it is convenient to rewrite the orbital angular momentum in terms of raising and lowering operators
\begin{equation}
L_{x} = \frac{1}{2}(L_{+}+L_{-}), \hspace{0.5cm} L_{y} = -\frac{i}{2}(L_{+}-L_{-})\;.
\end{equation}
The basis functions for $p_{x}$ and $p_{y}$ have the usual form 
\begin{eqnarray}
&&|p_{x}\rangle_{\uparrow} = \frac{\sqrt{2}}{2}(-|1, 1\rangle_{\uparrow}+|1, -1\rangle_{\uparrow})\;, \nonumber\\
&&|p_{y}\rangle_{\uparrow} = \frac{i\sqrt{2}}{2}(|1, 1\rangle_{\uparrow}+|1, -1\rangle_{\uparrow})\;.
\end{eqnarray}
It is straightforward to calculate the following matrix elements,
\begin{subequations}
\begin{align}
&\langle p_{y}|\vec{L}\cdot\vec{S}|p_{x}\rangle = i\sigma_{z}, \hspace{0.5cm} \langle p_{x}|\vec{L}\cdot\vec{S}|p_{y}\rangle = -i\sigma_{z}\;, \\
&\langle p_{z}|\vec{L}\cdot\vec{S}|p_{x}\rangle = -i\sigma_{y}, \hspace{0.5cm} \langle p_{x}|\vec{L}\cdot\vec{S}|p_{z}\rangle = i\sigma_{y}\;, \\
&\langle p_{z}|\vec{L}\cdot\vec{S}|p_{y}\rangle = i\sigma_{x}, \hspace{0.5cm} \langle p_{y}|\vec{L}\cdot\vec{S}|p_{z}\rangle = -i\sigma_{x}\;.
\end{align}
\end{subequations}

It is easy to see that, for the $\sigma$-bands,  there is an on-site intrinsic SOC arising from the $L_{z}\sigma_{z}$ term. 
After taking account this term, the Hamiltonian matrix for $\sigma$-bands becomes
\begin{equation}\label{Hss}
H^{\sigma\sigma}= 
\begin{pmatrix}
H^{\sigma\sigma}_{\uparrow\uparrow} & 0 \\
0 & H^{\sigma\sigma}_{\downarrow\downarrow}
\end{pmatrix}\;. 
\end{equation}
with $H^{\sigma\sigma}_{\uparrow\uparrow}$ and $H^{\sigma\sigma}_{\downarrow\downarrow}$ are given as 
\begin{eqnarray}
H^{\sigma\sigma}_{\uparrow\uparrow}&=&
\begin{pmatrix}
0 & -i\lambda_{so} & h_{xx}^{AB} & h_{xy}^{AB}\cr
\dagger & 0 & h_{yx}^{AB} & h_{yy}^{AB} \cr
\dagger & \dagger & 0 & -i\lambda_{so}\cr
\dagger & \dagger & \dagger & 0
\end{pmatrix}\;, \\
H^{\sigma\sigma}_{\downarrow\downarrow}&=&
\begin{pmatrix}
0 & i\lambda_{so} & h_{xx}^{AB} & h_{xy}^{AB}\cr
\dagger & 0 & h_{yx}^{AB} & h_{yy}^{AB} \cr
\dagger & \dagger & 0 & i\lambda_{so}\cr
\dagger & \dagger & \dagger & 0
\end{pmatrix}\;.
\end{eqnarray}
In contrast to the $L_{z}\sigma_{z}$ term, $L_{x}\sigma_{x}+L_{y}\sigma_{y}$ mixes the different spin sectors in $H^{\sigma\pi}$ and $H^{\pi\sigma}$, {\it i.e.}
\begin{equation}\label{Hmix}
H^{\sigma\pi}=(H^{\pi\sigma})^{\dagger} = 
\begin{pmatrix}
H^{\sigma\pi}_{\uparrow\uparrow} & H^{\sigma\pi}_{\uparrow\downarrow} \\
H^{\sigma\pi}_{\downarrow\uparrow} & H^{\sigma\pi}_{\downarrow\downarrow}
\end{pmatrix}\;,
\end{equation}
where $H^{\sigma\pi}_{\uparrow\uparrow}=H^{\sigma\pi}_{\downarrow\downarrow}$ is given in Eq.~(\ref{H_sp}) and $H^{\sigma\pi}_{\uparrow\downarrow}$, $(H^{\sigma\pi}_{\downarrow\uparrow})^{\dagger}$ are defined as
\begin{eqnarray}\label{Hsp}
H^{\sigma\pi}_{\uparrow\downarrow}&=&
\begin{pmatrix}
\lambda_{so} & 0 & 0 & 0 \\
-i\lambda_{so} & 0 & 0 & 0 \\
 0 & \lambda_{so} & 0 & 0 \\
 0 & -i\lambda_{so} & 0 & 0
\end{pmatrix}\\
H^{\sigma\pi}_{\downarrow\uparrow}&=&
\begin{pmatrix}
-\lambda_{so} & 0 & 0 & 0 \\
-i\lambda_{so} & 0 & 0 & 0 \\
 0 & -\lambda_{so} & 0 & 0 \\
 0 & -i\lambda_{so} & 0 & 0
\end{pmatrix}.
\end{eqnarray}

To account for the substrate-induced potential difference at the two sides of (Bi, Sb, As)/SiC systems , we introduce an effective electric field $\vec{E}$ with strength $\lambda_{E}$ along the $z$-direction. It couples the $s$ and $p_{z}$ orbitals in the same spin sector, 
\begin{equation}\label{Hpp}
H^{\pi\pi}= 
\begin{pmatrix}
H^{\pi\pi}_{\uparrow\uparrow} & 0 \\
0 & H^{\pi\pi}_{\downarrow\downarrow}
\end{pmatrix}\;,
\end{equation}
where the nonzero block $H^{\pi\pi}_{\uparrow\uparrow}=H^{\pi\pi}_{\downarrow\downarrow}$ is given as
\begin{equation}
\begin{pmatrix}
h_{zz}^{AA} & h_{zz}^{AB} & \lambda_{E} & 0 \cr
h_{zz}^{BA} & h_{zz}^{BB} & 0 & \lambda_{E} \cr
\lambda_{E} & 0 & 0 & h_{ss}^{AB} \cr
0 & \lambda_{E} & h_{ss}^{AB} & 0 
\end{pmatrix}
\end{equation}

Eqs.~(\ref{Hss}), (\ref{Hmix}), (\ref{Hsp}) and (\ref{Hpp}) consist of the Hamiltonian of (Bi, Sb, As)/SiC spanned on the complete basis of $s$ and $p$ orbitals. As only the $\sigma$-bands are of interest, we will apply second-order perturbation theory to effectively integrate out the $\pi$-bands, but keeping their effect on the low-energy sector of $\sigma$-bands, which gives rise to the low-energy effective Hamiltonian for $\sigma$-bands as
\begin{equation}
H_{eff}^{\sigma\sigma} \approx H^{\sigma\sigma} - H^{\sigma\pi}\cdot (H^{\pi\pi})^{-1} \cdot H^{\pi\sigma}\;.
\end{equation}
To further simply our calculation, only the on-site energy of $H^{\pi\pi}$ and the Stark-effect term will be considered (by setting $h_{zz}^{AB}=h_{zz}^{BA}=h_{ss}^{AB}=h_{ss}^{BA}=0$ in Eq.~(\ref{Hpp})). This will not qualitatively change our conclusion. 
Another simplification is to neglect the corrections to $H^{\sigma\sigma}$ in the same spin sector as they are smaller than that in Eq.~(\ref{Hss}). We will mainly consider the spin-mixed terms that completely arise from the coupling to the Bi $\pi$-bands. This leads to the Rashba-type SOC for the $\sigma$-bands. 

After some math, we get the following effective Hamiltonian for the $\sigma$-bands of (Bi, Sb, As)/SiC systems:
\begin{equation}\label{effective_H}
H_{eff}^{\sigma\sigma} = 
\begin{pmatrix}
H^{\sigma\sigma}_{\uparrow\uparrow} & H^{\sigma\sigma}_{\uparrow\downarrow} \cr
H^{\sigma\sigma}_{\downarrow\uparrow} & H^{\sigma\sigma}_{\downarrow\downarrow}
\end{pmatrix}
\mbox{with }
H^{\sigma\sigma}_{\uparrow\downarrow} = (H^{\sigma\sigma}_{\downarrow\uparrow})^{\dagger}=
\begin{pmatrix}
0 & 0 & a & b\cr
0 & 0 & b & c \cr
d & e & 0 & 0 \cr
e & f & 0 & 0 
\end{pmatrix}\;.
\end{equation}
Here the elements $a, b, c, d, e, f$ are given as
\begin{eqnarray}\label{Coefficient}
a&=&-2\sqrt{3}i\lambda_{R}e^{i\frac{\sqrt{3}}{2}k_{y}}\sin\frac{k_{x}}{2}\\
b&=&\lambda_{R}[1-e^{i\frac{\sqrt{3}}{2}k_{y}}(\cos\frac{k_{x}}{2}+\sqrt{3}\sin\frac{k_{x}}{2})] \\
c&=&-2i\lambda_{R}[1-e^{i\frac{\sqrt{3}}{2}k_{y}}\cos\frac{k_{x}}{2}] \\
d&=&-2\sqrt{3}i\lambda_{R}e^{-i\frac{\sqrt{3}}{2}k_{y}}\sin\frac{k_{x}}{2}\\
e&=&-\lambda_{R}[1-e^{-i\frac{\sqrt{3}}{2}k_{y}}(\cos\frac{k_{x}}{2}-\sqrt{3}\sin\frac{k_{x}}{2})] \\
f&=&2i\lambda_{R}[1-e^{-i\frac{\sqrt{3}}{2}k_{y}}\cos\frac{k_{x}}{2}]\;.
\end{eqnarray}
In the last step we have redefined the effective Rashba SOC $\lambda_{R}$ as $\lambda_{R}=\frac{\lambda_{so}}{\lambda_{E}}V_{sp\sigma}^{1}$. 

There are 4 free parameters in this effective model, {\it {\it i.e.},} $V_{pp\sigma}^{1}$, $V_{pp\pi}^{1}$, $\lambda_{so}$ and $\lambda_{R}$ which can be obtained by diagonializing the Hamiltonian at K-point and fitting the energy levels to the corresponding DFT (GGA) band structure. Table~\ref{table-fitting}  gives the fitting parameters for the three systems in unit of eV. 
\begin{table}[htbp]
\begin{tabular*}{0.45\textwidth}{@{\extracolsep{\fill}} |c|c|c|c|c|}
\hline
Systems & $V_{pp\sigma}^{1}$ (eV) & $V_{pp\pi}^{1}$ (eV) & $\lambda_{so}$ (eV) & $\lambda_{R}$ (eV) \cr 
\hline
Bi/SiC & 2.0 & -0.21 & 0.435 & 0.032 \cr
\hline
Sb/SiC & 2.0 & -0.11 & 0.2 & 0.015 \cr
\hline
As/SiC & 2.0 & -0.041 & 0.06 & 0.005 \cr
\hline
\end{tabular*}
\caption{Model parameters for Bi/SiC, Sb/SiC, and As/SiC systems.}
\label{table-fitting}
\end{table}


\begin{thebibliography}{37}%
\makeatletter
\providecommand \@ifxundefined [1]{%
 \@ifx{#1\undefined}
}%
\providecommand \@ifnum [1]{%
 \ifnum #1\expandafter \@firstoftwo
 \else \expandafter \@secondoftwo
 \fi
}%
\providecommand \@ifx [1]{%
 \ifx #1\expandafter \@firstoftwo
 \else \expandafter \@secondoftwo
 \fi
}%
\providecommand \natexlab [1]{#1}%
\providecommand \enquote  [1]{``#1''}%
\providecommand \bibnamefont  [1]{#1}%
\providecommand \bibfnamefont [1]{#1}%
\providecommand \citenamefont [1]{#1}%
\providecommand \href@noop [0]{\@secondoftwo}%
\providecommand \href [0]{\begingroup \@sanitize@url \@href}%
\providecommand \@href[1]{\@@startlink{#1}\@@href}%
\providecommand \@@href[1]{\endgroup#1\@@endlink}%
\providecommand \@sanitize@url [0]{\catcode `\\12\catcode `\$12\catcode
  `\&12\catcode `\#12\catcode `\^12\catcode `\_12\catcode `\%12\relax}%
\providecommand \@@startlink[1]{}%
\providecommand \@@endlink[0]{}%
\providecommand \url  [0]{\begingroup\@sanitize@url \@url }%
\providecommand \@url [1]{\endgroup\@href {#1}{\urlprefix }}%
\providecommand \urlprefix  [0]{URL }%
\providecommand \Eprint [0]{\href }%
\providecommand \doibase [0]{http://dx.doi.org/}%
\providecommand \selectlanguage [0]{\@gobble}%
\providecommand \bibinfo  [0]{\@secondoftwo}%
\providecommand \bibfield  [0]{\@secondoftwo}%
\providecommand \translation [1]{[#1]}%
\providecommand \BibitemOpen [0]{}%
\providecommand \bibitemStop [0]{}%
\providecommand \bibitemNoStop [0]{.\EOS\space}%
\providecommand \EOS [0]{\spacefactor3000\relax}%
\providecommand \BibitemShut  [1]{\csname bibitem#1\endcsname}%
\let\auto@bib@innerbib\@empty
\bibitem [{\citenamefont {K{\"o}nig}\ \emph {et~al.}(2007)\citenamefont
  {K{\"o}nig}, \citenamefont {Wiedmann}, \citenamefont {Br{\"u}ne},
  \citenamefont {Roth}, \citenamefont {Buhmann}, \citenamefont {Molenkamp},
  \citenamefont {Qi},\ and\ \citenamefont {Zhang}}]{Konig766}%
  \BibitemOpen
  \bibfield  {author} {\bibinfo {author} {\bibfnamefont {M.}~\bibnamefont
  {K{\"o}nig}}, \bibinfo {author} {\bibfnamefont {S.}~\bibnamefont {Wiedmann}},
  \bibinfo {author} {\bibfnamefont {C.}~\bibnamefont {Br{\"u}ne}}, \bibinfo
  {author} {\bibfnamefont {A.}~\bibnamefont {Roth}}, \bibinfo {author}
  {\bibfnamefont {H.}~\bibnamefont {Buhmann}}, \bibinfo {author} {\bibfnamefont
  {L.~W.}\ \bibnamefont {Molenkamp}}, \bibinfo {author} {\bibfnamefont {X.-L.}\
  \bibnamefont {Qi}}, \ and\ \bibinfo {author} {\bibfnamefont {S.-C.}\
  \bibnamefont {Zhang}},\ }\href {\doibase 10.1126/science.1148047} {\bibfield
  {journal} {\bibinfo  {journal} {Science}\ }\textbf {\bibinfo {volume}
  {318}},\ \bibinfo {pages} {766} (\bibinfo {year} {2007})}\BibitemShut
  {NoStop}%
\bibitem [{\citenamefont {Roth}\ \emph {et~al.}(2009)\citenamefont {Roth},
  \citenamefont {Br{\"u}ne}, \citenamefont {Buhmann}, \citenamefont
  {Molenkamp}, \citenamefont {Maciejko}, \citenamefont {Qi},\ and\
  \citenamefont {Zhang}}]{Roth294}%
  \BibitemOpen
  \bibfield  {author} {\bibinfo {author} {\bibfnamefont {A.}~\bibnamefont
  {Roth}}, \bibinfo {author} {\bibfnamefont {C.}~\bibnamefont {Br{\"u}ne}},
  \bibinfo {author} {\bibfnamefont {H.}~\bibnamefont {Buhmann}}, \bibinfo
  {author} {\bibfnamefont {L.~W.}\ \bibnamefont {Molenkamp}}, \bibinfo {author}
  {\bibfnamefont {J.}~\bibnamefont {Maciejko}}, \bibinfo {author}
  {\bibfnamefont {X.-L.}\ \bibnamefont {Qi}}, \ and\ \bibinfo {author}
  {\bibfnamefont {S.-C.}\ \bibnamefont {Zhang}},\ }\href {\doibase
  10.1126/science.1174736} {\bibfield  {journal} {\bibinfo  {journal}
  {Science}\ }\textbf {\bibinfo {volume} {325}},\ \bibinfo {pages} {294}
  (\bibinfo {year} {2009})}\BibitemShut {NoStop}%
\bibitem [{\citenamefont {Knez}\ \emph {et~al.}(2011)\citenamefont {Knez},
  \citenamefont {Du},\ and\ \citenamefont {Sullivan}}]{PhysRevLett.107.136603}%
  \BibitemOpen
  \bibfield  {author} {\bibinfo {author} {\bibfnamefont {I.}~\bibnamefont
  {Knez}}, \bibinfo {author} {\bibfnamefont {R.-R.}\ \bibnamefont {Du}}, \ and\
  \bibinfo {author} {\bibfnamefont {G.}~\bibnamefont {Sullivan}},\ }\href
  {\doibase 10.1103/PhysRevLett.107.136603} {\bibfield  {journal} {\bibinfo
  {journal} {Phys. Rev. Lett.}\ }\textbf {\bibinfo {volume} {107}},\ \bibinfo
  {pages} {136603} (\bibinfo {year} {2011})}\BibitemShut {NoStop}%
\bibitem [{\citenamefont {Hasan}\ and\ \citenamefont {Kane}(2010)}]{HasanKane}%
  \BibitemOpen
  \bibfield  {author} {\bibinfo {author} {\bibfnamefont {M.~Z.}\ \bibnamefont
  {Hasan}}\ and\ \bibinfo {author} {\bibfnamefont {C.~L.}\ \bibnamefont
  {Kane}},\ }\href@noop {} {\bibfield  {journal} {\bibinfo  {journal} {Rev.
  Mod. Phys.}\ }\textbf {\bibinfo {volume} {82}},\ \bibinfo {pages} {3045}
  (\bibinfo {year} {2010})}\BibitemShut {NoStop}%
\bibitem [{\citenamefont {Qi}\ and\ \citenamefont {Zhang}(2011)}]{xlreview}%
  \BibitemOpen
  \bibfield  {author} {\bibinfo {author} {\bibfnamefont {X.-L.}\ \bibnamefont
  {Qi}}\ and\ \bibinfo {author} {\bibfnamefont {S.-C.}\ \bibnamefont {Zhang}},\
  }\href@noop {} {\bibfield  {journal} {\bibinfo  {journal} {Rev. Mod. Phys.}\
  }\textbf {\bibinfo {volume} {83}},\ \bibinfo {pages} {1057} (\bibinfo {year}
  {2011})}\BibitemShut {NoStop}%
\bibitem [{\citenamefont {Moore}(2010)}]{Moore2010:N}%
  \BibitemOpen
  \bibfield  {author} {\bibinfo {author} {\bibfnamefont {J.~E.}\ \bibnamefont
  {Moore}},\ }\href@noop {} {\bibfield  {journal} {\bibinfo  {journal} {Nature
  (London)}\ }\textbf {\bibinfo {volume} {464}},\ \bibinfo {pages} {194}
  (\bibinfo {year} {2010})}\BibitemShut {NoStop}%
\bibitem [{\citenamefont {Kane}\ and\ \citenamefont
  {Mele}(2005{\natexlab{a}})}]{kane-05prl226801}%
  \BibitemOpen
  \bibfield  {author} {\bibinfo {author} {\bibfnamefont {C.~L.}\ \bibnamefont
  {Kane}}\ and\ \bibinfo {author} {\bibfnamefont {E.~J.}\ \bibnamefont
  {Mele}},\ }\href {\doibase 10.1103/PhysRevLett.95.226801} {\bibfield
  {journal} {\bibinfo  {journal} {Phys. Rev. Lett.}\ }\textbf {\bibinfo
  {volume} {95}},\ \bibinfo {pages} {226801} (\bibinfo {year}
  {2005}{\natexlab{a}})}\BibitemShut {NoStop}%
\bibitem [{\citenamefont {Bernevig}\ \emph {et~al.}(2006)\citenamefont
  {Bernevig}, \citenamefont {Hughes},\ and\ \citenamefont
  {Zhang}}]{Bernevig1757}%
  \BibitemOpen
  \bibfield  {author} {\bibinfo {author} {\bibfnamefont {B.~A.}\ \bibnamefont
  {Bernevig}}, \bibinfo {author} {\bibfnamefont {T.~L.}\ \bibnamefont
  {Hughes}}, \ and\ \bibinfo {author} {\bibfnamefont {S.-C.}\ \bibnamefont
  {Zhang}},\ }\href {\doibase 10.1126/science.1133734} {\bibfield  {journal}
  {\bibinfo  {journal} {Science}\ }\textbf {\bibinfo {volume} {314}},\ \bibinfo
  {pages} {1757} (\bibinfo {year} {2006})}\BibitemShut {NoStop}%
\bibitem [{\citenamefont {Wu}\ \emph {et~al.}(2006)\citenamefont {Wu},
  \citenamefont {Bernevig},\ and\ \citenamefont {Zhang}}]{Wu2006:PRL}%
  \BibitemOpen
  \bibfield  {author} {\bibinfo {author} {\bibfnamefont {C.}~\bibnamefont
  {Wu}}, \bibinfo {author} {\bibfnamefont {B.~A.}\ \bibnamefont {Bernevig}}, \
  and\ \bibinfo {author} {\bibfnamefont {S.-C.}\ \bibnamefont {Zhang}},\ }\href
  {\doibase 10.1103/PhysRevLett.96.106401} {\bibfield  {journal} {\bibinfo
  {journal} {Phys. Rev. Lett.}\ }\textbf {\bibinfo {volume} {96}},\ \bibinfo
  {pages} {106401} (\bibinfo {year} {2006})}\BibitemShut {NoStop}%
\bibitem [{\citenamefont {Xu}\ and\ \citenamefont {Moore}(2006)}]{Xu2006:PRB}%
  \BibitemOpen
  \bibfield  {author} {\bibinfo {author} {\bibfnamefont {C.}~\bibnamefont
  {Xu}}\ and\ \bibinfo {author} {\bibfnamefont {J.~E.}\ \bibnamefont {Moore}},\
  }\href {\doibase 10.1103/PhysRevB.73.045322} {\bibfield  {journal} {\bibinfo
  {journal} {Phys. Rev. B}\ }\textbf {\bibinfo {volume} {73}},\ \bibinfo
  {pages} {045322} (\bibinfo {year} {2006})}\BibitemShut {NoStop}%
\bibitem [{\citenamefont {Kane}\ and\ \citenamefont
  {Mele}(2005{\natexlab{b}})}]{PhysRevLett.95.146802}%
  \BibitemOpen
  \bibfield  {author} {\bibinfo {author} {\bibfnamefont {C.~L.}\ \bibnamefont
  {Kane}}\ and\ \bibinfo {author} {\bibfnamefont {E.~J.}\ \bibnamefont
  {Mele}},\ }\href {\doibase 10.1103/PhysRevLett.95.146802} {\bibfield
  {journal} {\bibinfo  {journal} {Phys. Rev. Lett.}\ }\textbf {\bibinfo
  {volume} {95}},\ \bibinfo {pages} {146802} (\bibinfo {year}
  {2005}{\natexlab{b}})}\BibitemShut {NoStop}%
\bibitem [{\citenamefont {V\"ayrynen}\ \emph {et~al.}(2013)\citenamefont
  {V\"ayrynen}, \citenamefont {Goldstein},\ and\ \citenamefont
  {Glazman}}]{PhysRevLett.110.216402}%
  \BibitemOpen
  \bibfield  {author} {\bibinfo {author} {\bibfnamefont {J.~I.}\ \bibnamefont
  {V\"ayrynen}}, \bibinfo {author} {\bibfnamefont {M.}~\bibnamefont
  {Goldstein}}, \ and\ \bibinfo {author} {\bibfnamefont {L.~I.}\ \bibnamefont
  {Glazman}},\ }\href {\doibase 10.1103/PhysRevLett.110.216402} {\bibfield
  {journal} {\bibinfo  {journal} {Phys. Rev. Lett.}\ }\textbf {\bibinfo
  {volume} {110}},\ \bibinfo {pages} {216402} (\bibinfo {year}
  {2013})}\BibitemShut {NoStop}%
\bibitem [{\citenamefont {K\"onig}\ \emph {et~al.}(2013)\citenamefont
  {K\"onig}, \citenamefont {Baenninger}, \citenamefont {Garcia}, \citenamefont
  {Harjee}, \citenamefont {Pruitt}, \citenamefont {Ames}, \citenamefont
  {Leubner}, \citenamefont {Br\"une}, \citenamefont {Buhmann}, \citenamefont
  {Molenkamp},\ and\ \citenamefont {Goldhaber-Gordon}}]{PhysRevX.3.021003}%
  \BibitemOpen
  \bibfield  {author} {\bibinfo {author} {\bibfnamefont {M.}~\bibnamefont
  {K\"onig}}, \bibinfo {author} {\bibfnamefont {M.}~\bibnamefont {Baenninger}},
  \bibinfo {author} {\bibfnamefont {A.~G.~F.}\ \bibnamefont {Garcia}}, \bibinfo
  {author} {\bibfnamefont {N.}~\bibnamefont {Harjee}}, \bibinfo {author}
  {\bibfnamefont {B.~L.}\ \bibnamefont {Pruitt}}, \bibinfo {author}
  {\bibfnamefont {C.}~\bibnamefont {Ames}}, \bibinfo {author} {\bibfnamefont
  {P.}~\bibnamefont {Leubner}}, \bibinfo {author} {\bibfnamefont
  {C.}~\bibnamefont {Br\"une}}, \bibinfo {author} {\bibfnamefont
  {H.}~\bibnamefont {Buhmann}}, \bibinfo {author} {\bibfnamefont {L.~W.}\
  \bibnamefont {Molenkamp}}, \ and\ \bibinfo {author} {\bibfnamefont
  {D.}~\bibnamefont {Goldhaber-Gordon}},\ }\href {\doibase
  10.1103/PhysRevX.3.021003} {\bibfield  {journal} {\bibinfo  {journal} {Phys.
  Rev. X}\ }\textbf {\bibinfo {volume} {3}},\ \bibinfo {pages} {021003}
  (\bibinfo {year} {2013})}\BibitemShut {NoStop}%
\bibitem [{\citenamefont {Reis}\ \emph {et~al.}(2017)\citenamefont {Reis},
  \citenamefont {Li}, \citenamefont {Dudy}, \citenamefont {Bauernfeind},
  \citenamefont {Glass}, \citenamefont {Hanke}, \citenamefont {Thomale},
  \citenamefont {Sch{\"a}fer},\ and\ \citenamefont {Claessen}}]{Reis2017:S}%
  \BibitemOpen
  \bibfield  {author} {\bibinfo {author} {\bibfnamefont {F.}~\bibnamefont
  {Reis}}, \bibinfo {author} {\bibfnamefont {G.}~\bibnamefont {Li}}, \bibinfo
  {author} {\bibfnamefont {L.}~\bibnamefont {Dudy}}, \bibinfo {author}
  {\bibfnamefont {M.}~\bibnamefont {Bauernfeind}}, \bibinfo {author}
  {\bibfnamefont {S.}~\bibnamefont {Glass}}, \bibinfo {author} {\bibfnamefont
  {W.}~\bibnamefont {Hanke}}, \bibinfo {author} {\bibfnamefont
  {R.}~\bibnamefont {Thomale}}, \bibinfo {author} {\bibfnamefont
  {J.}~\bibnamefont {Sch{\"a}fer}}, \ and\ \bibinfo {author} {\bibfnamefont
  {R.}~\bibnamefont {Claessen}},\ }\href {\doibase 10.1126/science.aai8142}
  {\bibfield  {journal} {\bibinfo  {journal} {Science}\ } (\bibinfo {year}
  {2017}),\ 10.1126/science.aai8142}\BibitemShut {NoStop}%
\bibitem [{\citenamefont {Hohenberg}\ and\ \citenamefont
  {Kohn}(1964)}]{PhysRev.136.B864}%
  \BibitemOpen
  \bibfield  {author} {\bibinfo {author} {\bibfnamefont {P.}~\bibnamefont
  {Hohenberg}}\ and\ \bibinfo {author} {\bibfnamefont {W.}~\bibnamefont
  {Kohn}},\ }\href {\doibase 10.1103/PhysRev.136.B864} {\bibfield  {journal}
  {\bibinfo  {journal} {Phys. Rev.}\ }\textbf {\bibinfo {volume} {136}},\
  \bibinfo {pages} {B864} (\bibinfo {year} {1964})}\BibitemShut {NoStop}%
\bibitem [{\citenamefont {Kohn}\ and\ \citenamefont
  {Sham}(1965)}]{PhysRev.140.A1133}%
  \BibitemOpen
  \bibfield  {author} {\bibinfo {author} {\bibfnamefont {W.}~\bibnamefont
  {Kohn}}\ and\ \bibinfo {author} {\bibfnamefont {L.~J.}\ \bibnamefont
  {Sham}},\ }\href {\doibase 10.1103/PhysRev.140.A1133} {\bibfield  {journal}
  {\bibinfo  {journal} {Phys. Rev.}\ }\textbf {\bibinfo {volume} {140}},\
  \bibinfo {pages} {A1133} (\bibinfo {year} {1965})}\BibitemShut {NoStop}%
\bibitem [{\citenamefont {Zhang}\ \emph {et~al.}(2014)\citenamefont {Zhang},
  \citenamefont {Li},\ and\ \citenamefont {Wu}}]{ZhangGF2014}%
  \BibitemOpen
  \bibfield  {author} {\bibinfo {author} {\bibfnamefont {G.-F.}\ \bibnamefont
  {Zhang}}, \bibinfo {author} {\bibfnamefont {Y.}~\bibnamefont {Li}}, \ and\
  \bibinfo {author} {\bibfnamefont {C.}~\bibnamefont {Wu}},\ }\href {\doibase
  10.1103/PhysRevB.90.075114} {\bibfield  {journal} {\bibinfo  {journal} {Phys.
  Rev. B}\ }\textbf {\bibinfo {volume} {90}},\ \bibinfo {pages} {075114}
  (\bibinfo {year} {2014})}\BibitemShut {NoStop}%
\bibitem [{\citenamefont {Wu}\ \emph {et~al.}(2007)\citenamefont {Wu},
  \citenamefont {Bergman}, \citenamefont {Balents},\ and\ \citenamefont
  {Das~Sarma}}]{WuCJ2007}%
  \BibitemOpen
  \bibfield  {author} {\bibinfo {author} {\bibfnamefont {C.}~\bibnamefont
  {Wu}}, \bibinfo {author} {\bibfnamefont {D.}~\bibnamefont {Bergman}},
  \bibinfo {author} {\bibfnamefont {L.}~\bibnamefont {Balents}}, \ and\
  \bibinfo {author} {\bibfnamefont {S.}~\bibnamefont {Das~Sarma}},\ }\href
  {\doibase 10.1103/PhysRevLett.99.070401} {\bibfield  {journal} {\bibinfo
  {journal} {Phys. Rev. Lett.}\ }\textbf {\bibinfo {volume} {99}},\ \bibinfo
  {pages} {070401} (\bibinfo {year} {2007})}\BibitemShut {NoStop}%
\bibitem [{\citenamefont {Wu}\ and\ \citenamefont
  {Das~Sarma}(2008)}]{WuCJ2008}%
  \BibitemOpen
  \bibfield  {author} {\bibinfo {author} {\bibfnamefont {C.}~\bibnamefont
  {Wu}}\ and\ \bibinfo {author} {\bibfnamefont {S.}~\bibnamefont {Das~Sarma}},\
  }\href {\doibase 10.1103/PhysRevB.77.235107} {\bibfield  {journal} {\bibinfo
  {journal} {Phys. Rev. B}\ }\textbf {\bibinfo {volume} {77}},\ \bibinfo
  {pages} {235107} (\bibinfo {year} {2008})}\BibitemShut {NoStop}%
\bibitem [{\citenamefont {Wu}(2008)}]{WuCJ2008a}%
  \BibitemOpen
  \bibfield  {author} {\bibinfo {author} {\bibfnamefont {C.}~\bibnamefont
  {Wu}},\ }\href {\doibase 10.1103/PhysRevLett.101.186807} {\bibfield
  {journal} {\bibinfo  {journal} {Phys. Rev. Lett.}\ }\textbf {\bibinfo
  {volume} {101}},\ \bibinfo {pages} {186807} (\bibinfo {year}
  {2008})}\BibitemShut {NoStop}%
\bibitem [{\citenamefont {Zhang}\ \emph {et~al.}(2011)\citenamefont {Zhang},
  \citenamefont {Hung}, \citenamefont {Zhang},\ and\ \citenamefont
  {Wu}}]{Zhang2011:PRA}%
  \BibitemOpen
  \bibfield  {author} {\bibinfo {author} {\bibfnamefont {M.}~\bibnamefont
  {Zhang}}, \bibinfo {author} {\bibfnamefont {H.-h.}\ \bibnamefont {Hung}},
  \bibinfo {author} {\bibfnamefont {C.}~\bibnamefont {Zhang}}, \ and\ \bibinfo
  {author} {\bibfnamefont {C.}~\bibnamefont {Wu}},\ }\href {\doibase
  10.1103/PhysRevA.83.023615} {\bibfield  {journal} {\bibinfo  {journal} {Phys.
  Rev. A}\ }\textbf {\bibinfo {volume} {83}},\ \bibinfo {pages} {023615}
  (\bibinfo {year} {2011})}\BibitemShut {NoStop}%
\bibitem [{\citenamefont {Kresse}\ and\ \citenamefont
  {Furthm{\"u}ller}(1996)}]{kresse1996}%
  \BibitemOpen
  \bibfield  {author} {\bibinfo {author} {\bibfnamefont {G.}~\bibnamefont
  {Kresse}}\ and\ \bibinfo {author} {\bibfnamefont {J.}~\bibnamefont
  {Furthm{\"u}ller}},\ }\href@noop {} {\bibfield  {journal} {\bibinfo
  {journal} {Phys. Rev. B}\ }\textbf {\bibinfo {volume} {54}},\ \bibinfo
  {pages} {11169} (\bibinfo {year} {1996})}\BibitemShut {NoStop}%
\bibitem [{\citenamefont {Hsu}\ \emph {et~al.}(2015)\citenamefont {Hsu},
  \citenamefont {Huang}, \citenamefont {Chuang}, \citenamefont {Kuo},
  \citenamefont {Liu}, \citenamefont {Lin},\ and\ \citenamefont
  {Bansil}}]{bansil}%
  \BibitemOpen
  \bibfield  {author} {\bibinfo {author} {\bibfnamefont {C.-H.}\ \bibnamefont
  {Hsu}}, \bibinfo {author} {\bibfnamefont {Z.-Q.}\ \bibnamefont {Huang}},
  \bibinfo {author} {\bibfnamefont {F.-C.}\ \bibnamefont {Chuang}}, \bibinfo
  {author} {\bibfnamefont {C.-C.}\ \bibnamefont {Kuo}}, \bibinfo {author}
  {\bibfnamefont {Y.-T.}\ \bibnamefont {Liu}}, \bibinfo {author} {\bibfnamefont
  {H.}~\bibnamefont {Lin}}, \ and\ \bibinfo {author} {\bibfnamefont
  {A.}~\bibnamefont {Bansil}},\ }\href
  {http://stacks.iop.org/1367-2630/17/i=2/a=025005} {\bibfield  {journal}
  {\bibinfo  {journal} {New Journal of Physics}\ }\textbf {\bibinfo {volume}
  {17}},\ \bibinfo {pages} {025005} (\bibinfo {year} {2015})}\BibitemShut
  {NoStop}%
\bibitem [{\citenamefont {Huang}\ \emph {et~al.}(2013)\citenamefont {Huang},
  \citenamefont {Chuang}, \citenamefont {Hsu}, \citenamefont {Liu},
  \citenamefont {Chang}, \citenamefont {Lin},\ and\ \citenamefont
  {Bansil}}]{Huang2013:PRB}%
  \BibitemOpen
  \bibfield  {author} {\bibinfo {author} {\bibfnamefont {Z.-Q.}\ \bibnamefont
  {Huang}}, \bibinfo {author} {\bibfnamefont {F.-C.}\ \bibnamefont {Chuang}},
  \bibinfo {author} {\bibfnamefont {C.-H.}\ \bibnamefont {Hsu}}, \bibinfo
  {author} {\bibfnamefont {Y.-T.}\ \bibnamefont {Liu}}, \bibinfo {author}
  {\bibfnamefont {H.-R.}\ \bibnamefont {Chang}}, \bibinfo {author}
  {\bibfnamefont {H.}~\bibnamefont {Lin}}, \ and\ \bibinfo {author}
  {\bibfnamefont {A.}~\bibnamefont {Bansil}},\ }\href {\doibase
  10.1103/PhysRevB.88.165301} {\bibfield  {journal} {\bibinfo  {journal} {Phys.
  Rev. B}\ }\textbf {\bibinfo {volume} {88}},\ \bibinfo {pages} {165301}
  (\bibinfo {year} {2013})}\BibitemShut {NoStop}%
\bibitem [{\citenamefont {Min}\ \emph {et~al.}(2006)\citenamefont {Min},
  \citenamefont {Hill}, \citenamefont {{Si}nitsyn}, \citenamefont {Sahu},
  \citenamefont {Kleinman},\ and\ \citenamefont
  {MacDonald}}]{PhysRevB.74.165310}%
  \BibitemOpen
  \bibfield  {author} {\bibinfo {author} {\bibfnamefont {H.}~\bibnamefont
  {Min}}, \bibinfo {author} {\bibfnamefont {J.~E.}\ \bibnamefont {Hill}},
  \bibinfo {author} {\bibfnamefont {N.~A.}\ \bibnamefont {{Si}nitsyn}},
  \bibinfo {author} {\bibfnamefont {B.~R.}\ \bibnamefont {Sahu}}, \bibinfo
  {author} {\bibfnamefont {L.}~\bibnamefont {Kleinman}}, \ and\ \bibinfo
  {author} {\bibfnamefont {A.~H.}\ \bibnamefont {MacDonald}},\ }\href {\doibase
  10.1103/PhysRevB.74.165310} {\bibfield  {journal} {\bibinfo  {journal} {Phys.
  Rev. B}\ }\textbf {\bibinfo {volume} {74}},\ \bibinfo {pages} {165310}
  (\bibinfo {year} {2006})}\BibitemShut {NoStop}%
\bibitem [{\citenamefont {Liu}\ \emph {et~al.}(2014)\citenamefont {Liu},
  \citenamefont {Guan}, \citenamefont {Song}, \citenamefont {Yang},
  \citenamefont {Yang},\ and\ \citenamefont {Yao}}]{PhysRevB.90.085431}%
  \BibitemOpen
  \bibfield  {author} {\bibinfo {author} {\bibfnamefont {C.-C.}\ \bibnamefont
  {Liu}}, \bibinfo {author} {\bibfnamefont {S.}~\bibnamefont {Guan}}, \bibinfo
  {author} {\bibfnamefont {Z.}~\bibnamefont {Song}}, \bibinfo {author}
  {\bibfnamefont {S.~A.}\ \bibnamefont {Yang}}, \bibinfo {author}
  {\bibfnamefont {J.}~\bibnamefont {Yang}}, \ and\ \bibinfo {author}
  {\bibfnamefont {Y.}~\bibnamefont {Yao}},\ }\href {\doibase
  10.1103/PhysRevB.90.085431} {\bibfield  {journal} {\bibinfo  {journal} {Phys.
  Rev. B}\ }\textbf {\bibinfo {volume} {90}},\ \bibinfo {pages} {085431}
  (\bibinfo {year} {2014})}\BibitemShut {NoStop}%
\bibitem [{\citenamefont {Zhou}\ \emph {et~al.}(2014)\citenamefont {Zhou},
  \citenamefont {Ming}, \citenamefont {Liu}, \citenamefont {Wang},
  \citenamefont {Li},\ and\ \citenamefont {Liu}}]{zhou}%
  \BibitemOpen
  \bibfield  {author} {\bibinfo {author} {\bibfnamefont {M.}~\bibnamefont
  {Zhou}}, \bibinfo {author} {\bibfnamefont {W.}~\bibnamefont {Ming}}, \bibinfo
  {author} {\bibfnamefont {Z.}~\bibnamefont {Liu}}, \bibinfo {author}
  {\bibfnamefont {Z.}~\bibnamefont {Wang}}, \bibinfo {author} {\bibfnamefont
  {P.}~\bibnamefont {Li}}, \ and\ \bibinfo {author} {\bibfnamefont
  {F.}~\bibnamefont {Liu}},\ }\href@noop {} {\bibfield  {journal} {\bibinfo
  {journal} {Proc. Natl. Acad. Sci.}\ }\textbf {\bibinfo {volume} {111}},\
  \bibinfo {pages} {14378} (\bibinfo {year} {2014})}\BibitemShut {NoStop}%
\bibitem [{\citenamefont {Dominguez}\ \emph {et~al.}(2018)\citenamefont
  {Dominguez}, \citenamefont {Scharf}, \citenamefont {Li}, \citenamefont
  {Sch\"afer}, \citenamefont {Claessen}, \citenamefont {Hanke}, \citenamefont
  {Thomale},\ and\ \citenamefont {Hankiewicz}}]{Dominguez2018}%
  \BibitemOpen
  \bibfield  {author} {\bibinfo {author} {\bibfnamefont {F.}~\bibnamefont
  {Dominguez}}, \bibinfo {author} {\bibfnamefont {B.}~\bibnamefont {Scharf}},
  \bibinfo {author} {\bibfnamefont {G.}~\bibnamefont {Li}}, \bibinfo {author}
  {\bibfnamefont {J.}~\bibnamefont {Sch\"afer}}, \bibinfo {author}
  {\bibfnamefont {R.}~\bibnamefont {Claessen}}, \bibinfo {author}
  {\bibfnamefont {W.}~\bibnamefont {Hanke}}, \bibinfo {author} {\bibfnamefont
  {R.}~\bibnamefont {Thomale}}, \ and\ \bibinfo {author} {\bibfnamefont
  {E.~M.}\ \bibnamefont {Hankiewicz}},\ }\href@noop {} {\bibfield  {journal}
  {\bibinfo  {journal} {arXiv:1803.02648}\ } (\bibinfo {year}
  {2018})}\BibitemShut {NoStop}%
\bibitem [{\citenamefont {Klitzing}\ \emph {et~al.}(1980)\citenamefont
  {Klitzing}, \citenamefont {Dorda},\ and\ \citenamefont
  {Pepper}}]{klitzing-80prl494}%
  \BibitemOpen
  \bibfield  {author} {\bibinfo {author} {\bibfnamefont {K.~v.}\ \bibnamefont
  {Klitzing}}, \bibinfo {author} {\bibfnamefont {G.}~\bibnamefont {Dorda}}, \
  and\ \bibinfo {author} {\bibfnamefont {M.}~\bibnamefont {Pepper}},\ }\href
  {\doibase 10.1103/PhysRevLett.45.494} {\bibfield  {journal} {\bibinfo
  {journal} {Phys. Rev. Lett.}\ }\textbf {\bibinfo {volume} {45}},\ \bibinfo
  {pages} {494} (\bibinfo {year} {1980})}\BibitemShut {NoStop}%
\bibitem [{\citenamefont {Thouless}\ \emph {et~al.}(1982)\citenamefont
  {Thouless}, \citenamefont {Kohmoto}, \citenamefont {Nightingale},\ and\
  \citenamefont {den Nijs}}]{thouless82-prl405}%
  \BibitemOpen
  \bibfield  {author} {\bibinfo {author} {\bibfnamefont {D.~J.}\ \bibnamefont
  {Thouless}}, \bibinfo {author} {\bibfnamefont {M.}~\bibnamefont {Kohmoto}},
  \bibinfo {author} {\bibfnamefont {M.~P.}\ \bibnamefont {Nightingale}}, \ and\
  \bibinfo {author} {\bibfnamefont {M.}~\bibnamefont {den Nijs}},\ }\href
  {\doibase 10.1103/PhysRevLett.49.405} {\bibfield  {journal} {\bibinfo
  {journal} {Phys. Rev. Lett.}\ }\textbf {\bibinfo {volume} {49}},\ \bibinfo
  {pages} {405} (\bibinfo {year} {1982})}\BibitemShut {NoStop}%
\bibitem [{\citenamefont {Haldane}(1988)}]{haldane88prl2015}%
  \BibitemOpen
  \bibfield  {author} {\bibinfo {author} {\bibfnamefont {F.~D.~M.}\
  \bibnamefont {Haldane}},\ }\href {\doibase 10.1103/PhysRevLett.61.2015}
  {\bibfield  {journal} {\bibinfo  {journal} {Phys. Rev. Lett.}\ }\textbf
  {\bibinfo {volume} {61}},\ \bibinfo {pages} {2015} (\bibinfo {year}
  {1988})}\BibitemShut {NoStop}%
\bibitem [{\citenamefont {Bl\"ochl}(1994)}]{PhysRevB.50.17953}%
  \BibitemOpen
  \bibfield  {author} {\bibinfo {author} {\bibfnamefont {P.~E.}\ \bibnamefont
  {Bl\"ochl}},\ }\href {\doibase 10.1103/PhysRevB.50.17953} {\bibfield
  {journal} {\bibinfo  {journal} {Phys. Rev. B}\ }\textbf {\bibinfo {volume}
  {50}},\ \bibinfo {pages} {17953} (\bibinfo {year} {1994})}\BibitemShut
  {NoStop}%
\bibitem [{\citenamefont {Perdew}\ \emph {et~al.}(1996)\citenamefont {Perdew},
  \citenamefont {Burke},\ and\ \citenamefont {Ernzerhof}}]{perdew1996}%
  \BibitemOpen
  \bibfield  {author} {\bibinfo {author} {\bibfnamefont {J.~P.}\ \bibnamefont
  {Perdew}}, \bibinfo {author} {\bibfnamefont {K.}~\bibnamefont {Burke}}, \
  and\ \bibinfo {author} {\bibfnamefont {M.}~\bibnamefont {Ernzerhof}},\
  }\href@noop {} {\bibfield  {journal} {\bibinfo  {journal} {Phys. Rev. Lett.}\
  }\textbf {\bibinfo {volume} {77}},\ \bibinfo {pages} {3865} (\bibinfo {year}
  {1996})}\BibitemShut {NoStop}%
\bibitem [{\citenamefont {Monkhorst}\ and\ \citenamefont
  {Pack}(1976)}]{PhysRevB.13.5188}%
  \BibitemOpen
  \bibfield  {author} {\bibinfo {author} {\bibfnamefont {H.~J.}\ \bibnamefont
  {Monkhorst}}\ and\ \bibinfo {author} {\bibfnamefont {J.~D.}\ \bibnamefont
  {Pack}},\ }\href {\doibase 10.1103/PhysRevB.13.5188} {\bibfield  {journal}
  {\bibinfo  {journal} {Phys. Rev. B}\ }\textbf {\bibinfo {volume} {13}},\
  \bibinfo {pages} {5188} (\bibinfo {year} {1976})}\BibitemShut {NoStop}%
\bibitem [{\citenamefont {Marzari}\ and\ \citenamefont
  {Vanderbilt}(1997)}]{Marzari1997}%
  \BibitemOpen
  \bibfield  {author} {\bibinfo {author} {\bibfnamefont {N.}~\bibnamefont
  {Marzari}}\ and\ \bibinfo {author} {\bibfnamefont {D.}~\bibnamefont
  {Vanderbilt}},\ }\href {\doibase 10.1103/PhysRevB.56.12847} {\bibfield
  {journal} {\bibinfo  {journal} {Phys. Rev. B}\ }\textbf {\bibinfo {volume}
  {56}},\ \bibinfo {pages} {12847} (\bibinfo {year} {1997})}\BibitemShut
  {NoStop}%
\bibitem [{\citenamefont {Mostofi}\ \emph {et~al.}(2008)\citenamefont
  {Mostofi}, \citenamefont {Yates}, \citenamefont {Lee}, \citenamefont {Souza},
  \citenamefont {Vanderbilt},\ and\ \citenamefont {Marzari}}]{Mostofi2008}%
  \BibitemOpen
  \bibfield  {author} {\bibinfo {author} {\bibfnamefont {A.~A.}\ \bibnamefont
  {Mostofi}}, \bibinfo {author} {\bibfnamefont {J.~R.}\ \bibnamefont {Yates}},
  \bibinfo {author} {\bibfnamefont {Y.-S.}\ \bibnamefont {Lee}}, \bibinfo
  {author} {\bibfnamefont {I.}~\bibnamefont {Souza}}, \bibinfo {author}
  {\bibfnamefont {D.}~\bibnamefont {Vanderbilt}}, \ and\ \bibinfo {author}
  {\bibfnamefont {N.}~\bibnamefont {Marzari}},\ }\href@noop {} {\bibfield
  {journal} {\bibinfo  {journal} {Comput. Phys. Commun.}\ }\textbf {\bibinfo
  {volume} {178}},\ \bibinfo {pages} {685} (\bibinfo {year}
  {2008})}\BibitemShut {NoStop}%
\bibitem [{\citenamefont {Soluyanov}\ and\ \citenamefont
  {Vanderbilt}(2011)}]{Vanderbilt2011}%
  \BibitemOpen
  \bibfield  {author} {\bibinfo {author} {\bibfnamefont {A.~A.}\ \bibnamefont
  {Soluyanov}}\ and\ \bibinfo {author} {\bibfnamefont {D.}~\bibnamefont
  {Vanderbilt}},\ }\href {\doibase 10.1103/PhysRevB.83.235401} {\bibfield
  {journal} {\bibinfo  {journal} {Phys. Rev. B}\ }\textbf {\bibinfo {volume}
  {83}},\ \bibinfo {pages} {235401} (\bibinfo {year} {2011})}\BibitemShut
  {NoStop}%
\end{thebibliography}

%

\end{document}